\documentclass[%
 reprint,
usenatbib,
 amsmath,amssymb,
 aps,
 nofootinbib, 
twocolumn]{aastex631}

\usepackage{graphicx}
\usepackage{dcolumn}
\usepackage{tabularx}
\usepackage{bm}

\usepackage{booktabs, tabularx, makecell, multirow, xspace}
\usepackage{graphicx}	
\usepackage{amsmath}	
\usepackage{amssymb}	
\usepackage{physics}   
\usepackage{comment}   
\usepackage{slashed}

\newcommand{\msun}{{\,\rm M_\odot}}

\newcommand{\kms}{\,{\rm km}\,{\rm s}^{-1}}
\newcommand{\cm}{\,{\rm cm}}

\newcommand{\K}{\,{\rm K}}

\newcommand{\pc}{\,{\rm pc}}
\newcommand{\kpc}{\,{\rm kpc}}

\newcommand{\tIE}{t_{\rm IE}}

\newcommand{\me}{m_{\rm e}}
\newcommand{\mproton}{m_{\rm p}}
\renewcommand{\arraystretch}{1.8}
\newcolumntype{C}[1]{>{\centering\let\newline\\\arraybackslash\hspace{0pt}}m{#1}}

\definecolor{nicered}{rgb}{0.7,0.1,0.1}

\def\aap{A\&A}

\def\apjl{ApJ}

\def\apjs{ApJS}

\newcommand{\phat}{p'}
\newcommand{\ehat}{e'}
\newcommand{\alphahat}{\alpha'}
\newcommand{\mehat}{m_{e'}}
\newcommand{\mphat}{m_{p'}}

\newcommand{\Hhat}{H'}

\newcommand{\mstar}{M_\odot}

\newcommand{\nehat}{n_{\ehat}}
\newcommand{\nphat}{n_{\phat}}
\newcommand{\nHhat}{n_{\scriptscriptstyle{\Hhat}}}
\newcommand{\nHneutralhat}{n_{\scriptscriptstyle{H_0'}}}
\newcommand{\xehat}{x_{\ehat}}
\newcommand{\xphat}{x_{\phat}}
\newcommand{\xHneutralhat}{x_{\scriptscriptstyle{H_0'}}}

\newcommand{\tff}{t_{\rm ff}}

\newcommand{\zhalf}{z_{1/2}}
\newcommand{\rhalf}{r_{1/2}}

\newcommand{\fgas}{f_{\rm gas}}
\newcommand{\cdm}{\texttt{CDM}}
\newcommand{\cdmnf}{\texttt{CDM-NF}}
\newcommand{\admone}{\texttt{ADM-1}}
\newcommand{\admtwo}{\texttt{ADM-2}}

\newcommand{\fthin}{f_{\rm thin}}
\newcommand{\fthick}{f_{\rm thick}}

\newcommand{\fspheroid}{f_{\rm spheroid}}

\newcommand{\zninety}{z_{9/10}}
\newcommand{\flatness}{\Tilde{f}}

\begin{document}


\title{Simulating Atomic Dark Matter in Milky Way Analogues}

\author[0000-0002-7638-7454]{Sandip Roy}
\affiliation{Department of Physics, Princeton University, Princeton, NJ 08544, USA}

\author[0000-0002-6196-823X]{Xuejian Shen}
\affiliation{TAPIR, California Institute of Technology, Pasadena, CA, 91125, USA}

\author[0000-0002-8495-8659]{Mariangela Lisanti}
\affiliation{Department of Physics, Princeton University, Princeton, NJ 08544, USA}
\affiliation{Center for Computational Astrophysics, Flatiron Institute, New York, NY 10010, USA}

\author[0000-0003-0263-6195]{David Curtin}
\affiliation{Department of Physics, University of Toronto, Toronto, Ontario M5S 1A7, Canada}

\author[0000-0002-8659-3729]{Norman Murray}
\affiliation{Canadian Institute for Theoretical Astrophysics, University of Toronto, Toronto, ON M5S 3H8, Canada}

\author[0000-0003-3729-1684]{Philip F. Hopkins}
\affiliation{TAPIR, California Institute of Technology, Pasadena, CA, 91125, USA}

\begin{abstract}

Dark sector theories naturally lead to multi-component scenarios for dark matter where a sub-component can dissipate energy through self-interactions, allowing it to efficiently cool inside galaxies. 
We present the first cosmological hydrodynamical simulations of Milky Way analogues where the majority of dark matter is collisionless Cold Dark Matter~(CDM), but a sub-component (6\%) is strongly dissipative minimal Atomic Dark Matter~(ADM). The simulations, implemented in \texttt{GIZMO} and utilizing FIRE-2 galaxy formation physics to model the standard baryonic sector, demonstrate that the addition of even a small fraction of dissipative dark matter can significantly impact galactic evolution despite being consistent with current cosmological constraints. 
We show that ADM gas with roughly Standard-Model-like masses and couplings can cool to form a rotating ``dark disk'' with angular momentum closely aligned with the visible stellar disk.  The morphology of the disk depends sensitively on the parameters of the ADM model, which affect the cooling rates in the dark sector. 
The majority of the ADM gas gravitationally collapses into dark ``clumps’’ (regions of black hole or mirror star formation), which form a prominent bulge and a rotating thick disk in the central galaxy.  These clumps form early and quickly sink to the inner $\sim$ kpc of the galaxy, affecting the galaxy's star-formation history and present-day baryonic and CDM distributions. 
\end{abstract}



\section{\label{sec:intro}Introduction}

The Cold Dark Matter~(CDM) model successfully describes structure on the largest scales of the Universe~\citep{Planck:2018vyg}. 
However, strong theoretical arguments motivate extensions to the CDM framework that alter structure on sub-galactic scales: a more complex ``dark sector'' may contain multiple DM particles and new forces~\citep{Bertone:2018krk}. These include  Hidden Valleys~\citep{Strassler:2006im}  
and theories where the dark sector is related to the Standard Model by a discrete symmetry, such as the Twin Higgs model~\citep{Chacko:2005pe, Chacko:2016hvu, Chacko:2018vss},
which solves the electroweak hierarchy problem, 
and Mirror DM~\citep{Berezhiani:2005vv,
Foot:2004pa,
Mohapatra:2000qx}.

Dark sectors frequently include a fraction of DM with dissipative self-interactions, which can cool and clump in opposition to the purely gravitational dynamics of CDM. 
A useful benchmark model for such scenarios is  Atomic Dark Matter~(ADM), which consists of a dark proton, $\phat$, and dark electron, $\ehat$, that interact through a massless dark photon with coupling, $\alphahat$~\citep{Kaplan2009}.  The ADM can form a dark hydrogen bound-state and radiatively cools in direct analogy to the Standard Model. 
We  assume a minimal model that has no dark nuclear physics and only couples to the Standard Model through gravity. Moreover, we assume that the ADM abundance is set asymmetrically, so that the abundance of dark anti-particles is negligible~\citep{Zurek2013,Kaplan2011}. 
Minimal ADM is a plausible theory in its own right, but also arises within the more complete 
 frameworks mentioned above, making it a highly motivated subject for detailed study.  

 \begin{table*}[t]
\footnotesize
\begin{center}
\renewcommand{\arraystretch}{1.5}
\begin{tabular}{c|c|cccccccccc}
  \Xhline{3\arrayrulewidth}
\textbf{Simulation}&\textbf{Included Species} &$\mathbf{\frac{\Omega_{\rm cdm}}{\Omega_{\rm m}}}$ & $\mathbf{\frac{{m_{\rm cdm}}}{{M_\odot}}}$ & $\mathbf{\frac{\Omega_{\rm adm}}{\Omega_{\rm dm}}}$ &  $\mathbf{\frac{m_{\rm adm}}{M_\odot}}$ & $\mathbf{\frac{\Omega_{\rm b}}{\Omega_{\rm m}}}$ & $\mathbf{\frac{m_{\rm b}}{M_\odot}}$ & $\boldsymbol{\frac{\alpha'}{\alpha}}$ & $\mathbf{\frac{m_{p'}}{m_p}}$& $\mathbf{\frac{m_{e'}}{m_e}}$ & $\mathbf{\frac{T_{\rm cmb'}}{T_{\rm cmb}}}$\\
\hline
$\cdm$ & CDM+Bar. & 0.83 & 2.79$\times 10^{5}$ & 0 & - & 0.17 & $5.6\times10^4$ & - & - & - & -   \\
$\cdmnf$ & CDM+Bar., no FB & 0.83 & 2.79$\times 10^{5}$ & 0 & - & 0.17 &  $5.6\times10^4$ &- & - & - & -   \\
$\admone$ & CDM+ADM-1+Bar.  & 0.78 & 2.62$\times 10^{5}$ & 0.06 &  1.67$\times 10^{4}$ & 0.17 & $5.6\times10^4$ &$1/\sqrt{0.55}$ & $1.3$ & $0.55$ & 0.39  \\
$\admtwo$ &CDM+ADM-2+Bar.  & 0.78 & 2.62$\times 10^{5}$ & 0.06 &  1.67$\times 10^{4}$ & 0.17 & $5.6\times10^4$ &$2.5$ & $1.3$ & $0.55$ & 0.39  \\
  \Xhline{3\arrayrulewidth}
\end{tabular}
\end{center}
\caption{\label{tab:ADMspecies} A summary of the dark matter properties for the simulations studied in this work. Two simulations without ADM (with and without baryonic feedback) are compared to two simulations that include ADM. 
In all cases, the total DM abundance is $\Omega_\mathrm{dm} = 0.83$ with the remaining matter assumed to be baryons~(Bar.).  
The ADM, when present, constitutes 6\% of the total DM, with its cooling physics determined by its fine-structure constant ($\alpha'$), proton mass ($m_{p'}$), electron mass ($m_{e'}$) and dark CMB temperature $T_\mathrm{cmb'}$ relative to the Standard Model (unprimed quantities). 
 $m_\mathrm{cdm, adm, b}$ are the particle masses for different components in the simulation.
}
\end{table*}

On cosmic scales, ADM manifests through dark acoustic oscillations and contributions to $\Delta N_\mathrm{eff}$~\citep{Cyr-Racine:2013fsa, Gurian:2021qhk}. Current cosmological constraints allow for ADM to comprise $\gtrsim \mathcal{O}(10\%)$ of the DM for a wide range of parameters~\citep{Bansal:2021dfh, Bansal:2022qbi}. On much smaller scales, the signatures of an ADM sub-component can potentially be spectacular. ADM gas clouds can collapse and condense into dark compact objects, giving rise to 
dark white dwarfs~\citep{Ryan:2022hku}
and non-stellar-mass black holes~\citep{Shandera:2018xkn, Fernandez:2022zmc},  as well as 
mirror (neutron) stars~\citep{Curtin:2019ngc, Curtin:2019lhm, Hippert:2021fch, Hippert:2022snq}
(if there is dark nuclear physics). 

On galactic scales, the effects of a dissipative DM subcomponent could be equally dramatic, including the formation of a dark disk and modifications to halo structure~\citep{Fan:2013yva, Ghalsasi2017, Chang2018, Huo2020, Shen2021, Shen2022, Ryan:2021tgw}. This behavior is remarkably different from CDM, which survives in an extended halo because it experiences no energy loss. 
However, a quantitative understanding of ADM on galactic scales is almost completely lacking, since the formation of galactic structures and their time evolution depend on non-linear gravitational interactions involving the ADM, CDM and baryons. 
This Letter presents the first cosmological hydrodynamical simulations of Milky Way analogs that include ADM, opening a window into the detailed galactic physics of strongly dissipative DM.

To build intuition, Sec.~\ref{sec:basics} introduces the cooling physics of ADM by comparison to baryons.  Sec.~\ref{sec:sim_setup} describes the development of an ADM module for the \texttt{GIZMO} multi-physics code~\citep{Hopkins2015}, which is used to perform Milky Way zoom-ins of a cosmology where ADM comprises 6\% of DM.
Sec.~\ref{sec:sim_results} demonstrates that, for the two benchmarks studied here, the presence of a strongly cooling ADM sub-component leads to the formation of a centrally rotating ADM disk in the host galaxy.  However, the majority of the ADM exists in dense concentrations of collapsed gas (referred to as ``clumps''), which form early and rapidly accumulate in the inner $\sim\text{kpc}$ of the host. The presence of the ADM can dramatically affect the distribution of the baryonic gas and stars in the inner galaxy, even though it comprises only a small fraction of the total DM. These results are supplemented by several appendices.  Appendices~\ref{sec:dark_cooling_rates}--\ref{sec:gizmooverview} detail the cooling physics and assumptions made in the simulations, while App.~\ref{sec:supplementary_figures} contains supplementary figures and tables.
\\ 

\section{ADM Cooling Physics}
\label{sec:basics}

We simulate two ADM benchmarks that are consistent with cosmological bounds~\citep{Bansal:2022qbi}, but near the boundary of constrained parameter space, making additional astrophysical probes especially interesting. 
These benchmarks, described in Tab.~\ref{tab:ADMspecies}, feature Standard-Model-like masses and couplings, but still produce remarkably different outcomes. 

\begin{figure}
    \centering 
    \includegraphics[trim={0cm 0cm 0cm 0cm},clip,width=0.5\textwidth]{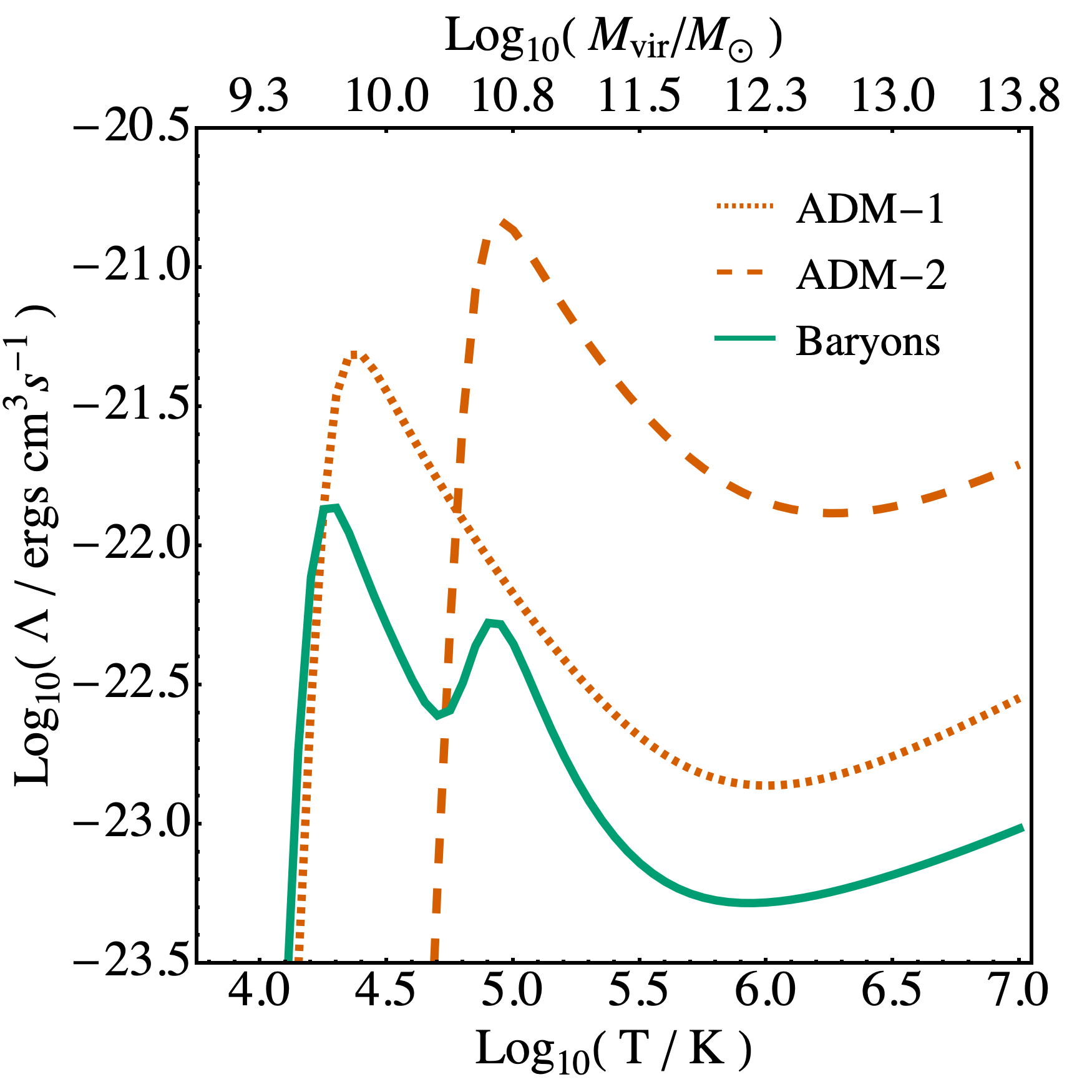}
    \caption{\label{fig:cooling_curves} Atomic cooling rates for baryons~(green line) and the ADM benchmarks, ADM-1 and ADM-2, which are defined in Tab.~\ref{tab:ADMspecies}~(dotted and dashed orange lines, respectively). The upper axis translates the temperature $T$ to the virial temperature of a galaxy with virial mass $M_{\rm vir}$ \citep{Bryan1997}. Rates are normalized by the particle number density to separate the effect of absolute density and mass on the shapes of the curves. The ADM cooling rates increase with decreasing dark electron mass and increasing dark fine-structure coupling. The cut-off temperature scales like the hydrogen binding energy. Compared to ADM-1, ADM-2 gas accretes more rapidly into the inner galaxy but is also more resistant to fragmentation due to its larger binding energy.}
\end{figure}

Fig.~\ref{fig:cooling_curves} displays volumetric cooling rates for dark hydrogen and  baryons, as a function of  temperature: $\Lambda = ({\rm d}E/{\rm d}t\,{\rm d}V)/n^2$, with ${\rm d}E/{\rm d}t\,{\rm d}V$ the rate of energy loss per unit volume and $n$ the species number density. All cooling curves  include collisional ionisation and excitation, recombination, and Bremsstrahlung processes. ADM cooling is modeled using the equations derived in~\citet{Rosenberg2017}, assuming ionisation-recombination equilibrium (App.~\ref{sec:dark_cooling_rates}). Dark molecular cooling~\citep{Ryan:2021dis} is not considered in this work as it affects dense regions of ADM gas below the resolution of our simulations, but can be added in the  future~\citep{Ryan:2021tgw}.

The baryonic cooling curve is distinguished by two peaks above the hydrogen binding energy $\sim10^4$~K: the first arises primarily from  collisional excitation of hydrogen and the second from collisional excitation of helium. Above $\sim10^6$~K, the baryonic gas is largely ionized and Bremsstrahlung dominates. 

There are three key differences between the baryonic and ADM cooling curves. First, the ADM curves do not have the characteristic second peak observed at low temperatures in the baryons because there is no dark helium present. Second, the volumetric cooling rate for each ADM model is faster than that for baryons. This is a function of the chosen $\ehat$ mass and $\alphahat$ coupling, and it governs how quickly a species collapses towards the central galaxy as well as the properties of the structures that it ultimately forms. We give the ADM greater volumetric cooling rates so that it achieves similar cooling timsecales to baryons, despite  having a lower mass fraction.  Lastly, differences in the binding energy of the (dark) hydrogen shift the cut-off temperature of each species, which dictates when atomic cooling effectively stops. This, in turn, dictates the minimum sound speed of the central (ADM) gas and its ability to gravitationally collapse, with greater cut-off temperatures corresponding to greater densities that have to be reached before loss of pressure support.  

\section{Simulations}
\label{sec:sim_setup}

Our suite includes four cosmological hydrodynamical zoom-in simulations of Milky Way analogues~(Tab.~\ref{tab:ADMspecies}). For a baseline comparison, we include two simulations where the matter component is comprised of only baryons and CDM.  Referenced as $\cdm$ and $\cdmnf$ , the former includes stellar feedback while the latter does not.  We also run two simulations for ADM ($\admone$ and $\admtwo$) that each include CDM and baryons with feedback. 

All simulations utilize~\texttt{GIZMO}~\citep{Hopkins2015}, with hydrodynamics solved using the mesh-free Lagrangian Godunov ``MFM'' method and gravity solved with an improved version of the Tree--PM solver~\citep{Springel2005}. Baryons are modeled using the hydrodynamics and galaxy formation physics from the Feedback In Realistic Environments~(FIRE) project~\citep{Hopkins2014}, specifically the ``FIRE-2'' version~\citep{Hopkins2017b,Hopkins2018}. The baryons include gas cooling down to the molecular phase ($\sim10$--$10^{10}\,{\rm K}$)~\citep{Ferland1998-CLOUDY, Wiersma2009}, heating from a meta-galactic radiation background~\citep{Onorbe2016,FG2020} and stellar sources, star formation, as well as explicit models for stellar and supernovae feedback \citep{Hopkins2014}. For $\cdmnf$, we remove stellar and supernovae feedback, as well as radiative feedback. The baryonic gas particles in both CDM simulations are converted into collisionless star particles once the gas reaches a density greater than $\sim1000~\cm^{-3}$, hosts non-zero molecular fractions \citep{Krumholz2011} and becomes Jeans-unstable and locally self-gravitating \citep{Hopkins2013}. 

ADM is implemented in \texttt{GIZMO} as a separate gas species that is completely decoupled from the baryonic gas, except for gravity. The dynamics of the ADM gas are solved with the same quasi-Lagrangian method used for baryons. The ADM cooling functions are implemented as described in App.~\ref{sec:dark_cooling_rates}, assuming no meta-galactic dark radiation background in the ADM sector except for the dark CMB. ADM gas cells that are locally self-gravitating and Jeans-unstable are turned into ``clump'' particles over the free-fall time scale. These dense clumps are expected to collapse in a run-away fashion to compact objects that effectively behave as collisionless particles at the resolved scale of the simulations, see e.g.~\citet{Gurian:2022nbx}. The clump formation criteria for ADM gas are different from the ones used for stellar particles because baryons have further constraints on molecular gas fractions and a set density threshold. We underscore that the ADM gas has no self-driven feedback, experiences no baryonic feedback, and does not undergo molecular cooling. For the ADM gas in \texttt{ADM-2}, after $z\approx 0.8$, we additionally allow clump formation if they have physical softening lengths $h^{\rm adm}_{\rm gas} \leq 7\pc$ and temperatures $T<10^{4.5}\K$ to prevent isolated dark ``molecular'' clouds from slowing the run time. 

The force softening for both the baryonic and ADM gas particles uses the fully conservative adaptive algorithm from~\citet{Price2007}, where the gravitational force assumes the identical mass distribution as the hydrodynamic equations.
The minimum gas softening reached is $h_{\rm gas} = 1.4\pc$. The CDM force resolution of the simulations is set to $h_{\rm dm} = 40\pc$. 

The main target halo(s) in the zoom-in region are picked from the standard FIRE-2 suite as described in~\citet{Wetzel2022,Hopkins2014}. This Letter focuses specifically on the Milky Way-like halo \texttt{m12i}, which has a total CDM mass of $\sim 10^{12}\msun$, a stellar disk of mass $7.4\times10^{10}\msun$, and a quiet merger history below $z < 1$ for  $\cdm$~\citep{Hopkins2018}. The initial transfer functions for the baryons, CDM, and ADM particles are calculated using a modified version of \texttt{CLASS}~\citep{Blas2011,Bansal:2021dfh,Bansal:2022qbi}. \texttt{MUSIC}~\citep{Hahn2011} is then used to generate initial conditions at $z\sim 100$ for the requisite mixture of baryons, CDM, and ADM.  The cosmological parameters that must be specified include the dark Cosmic Microwave Background~(CMB) temperature, $T_{\rm cmb'}$, and the ADM mass fraction.

\section{Galaxy Evolution in ADM}
\label{sec:sim_results}

\begin{figure*}
    \centering
    \includegraphics[width=0.49\linewidth]{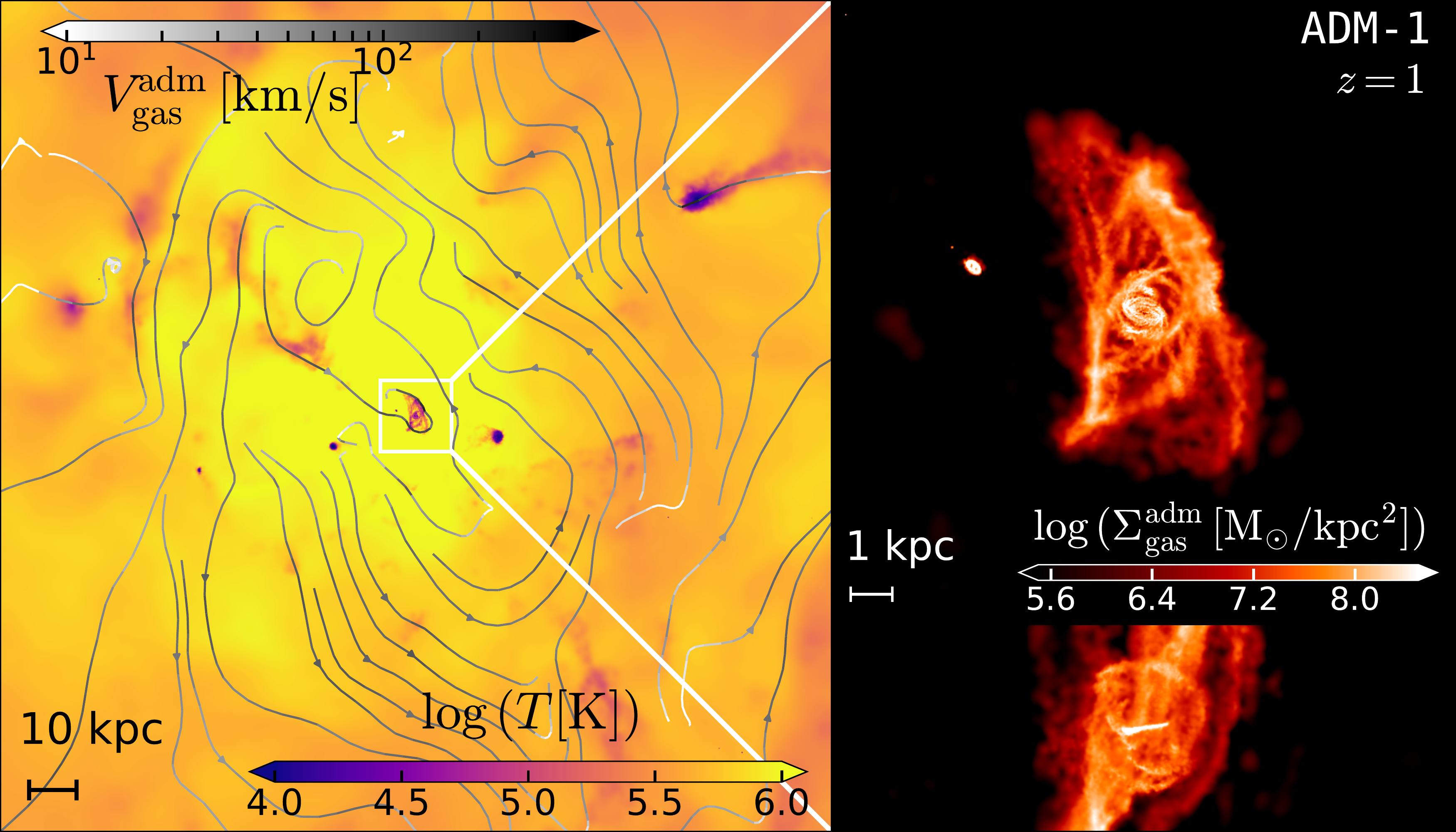}
    \includegraphics[width=0.49\linewidth]{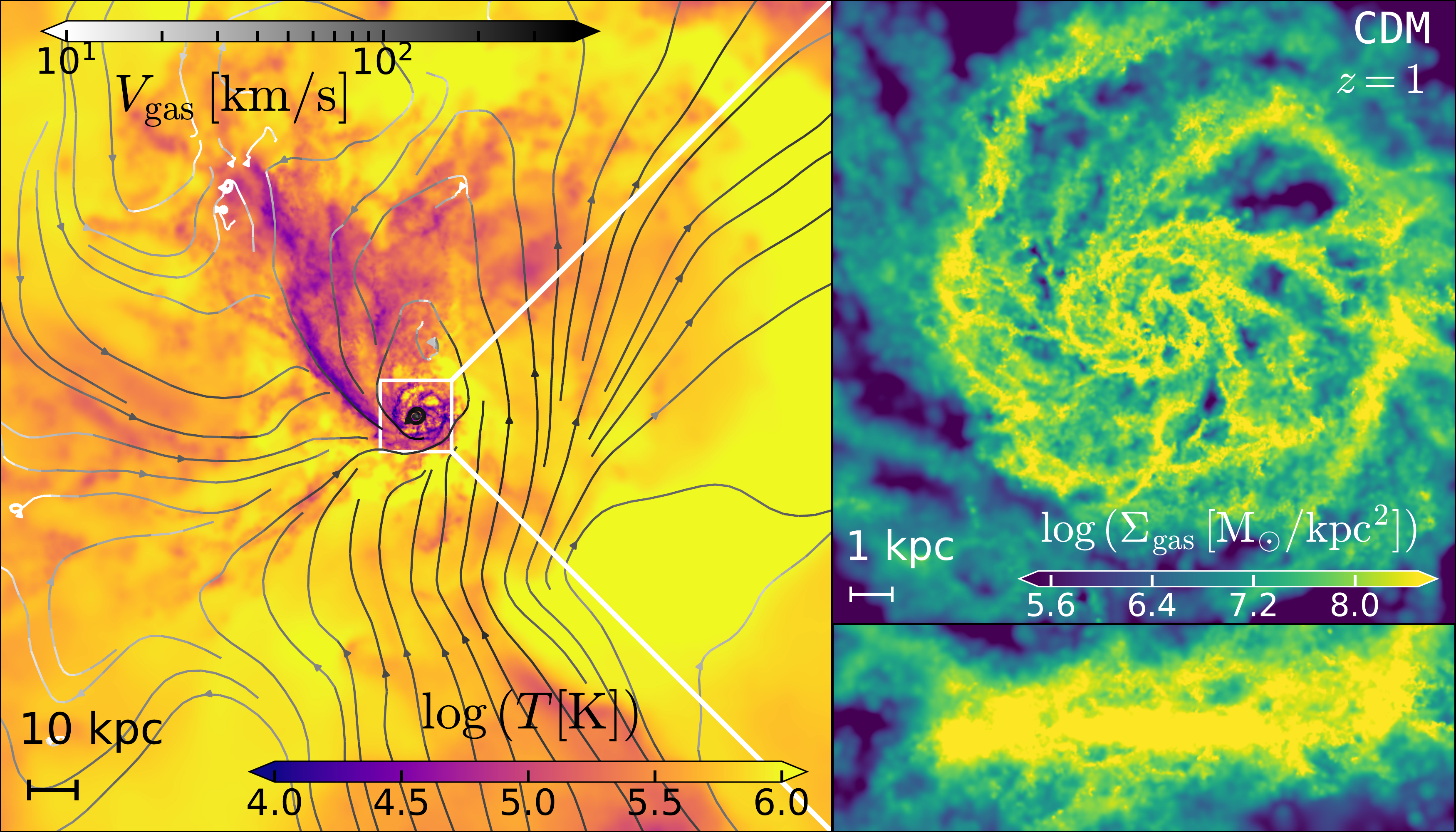}
    \includegraphics[width=0.49\linewidth]{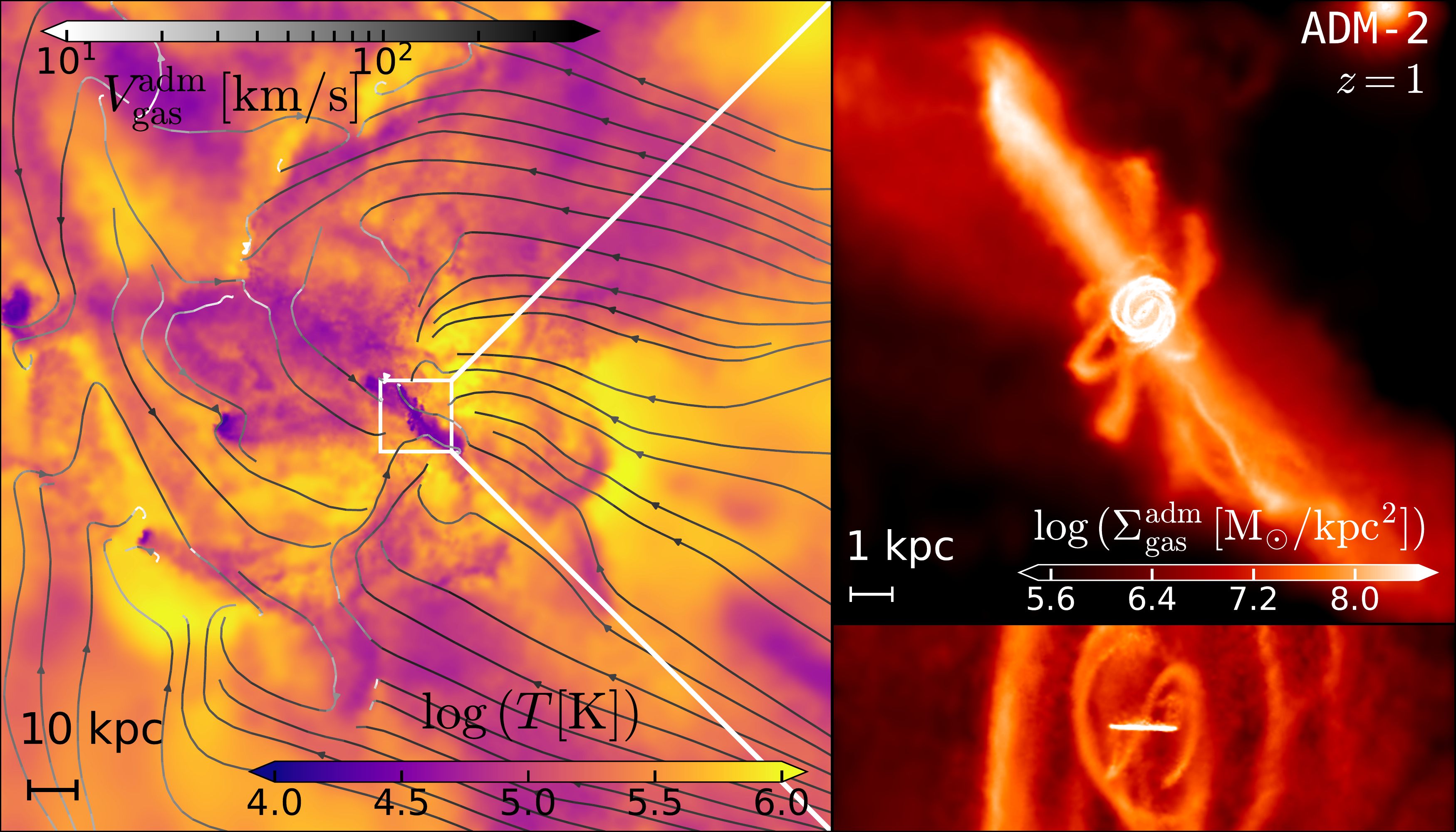}
    \includegraphics[width=0.49\linewidth]{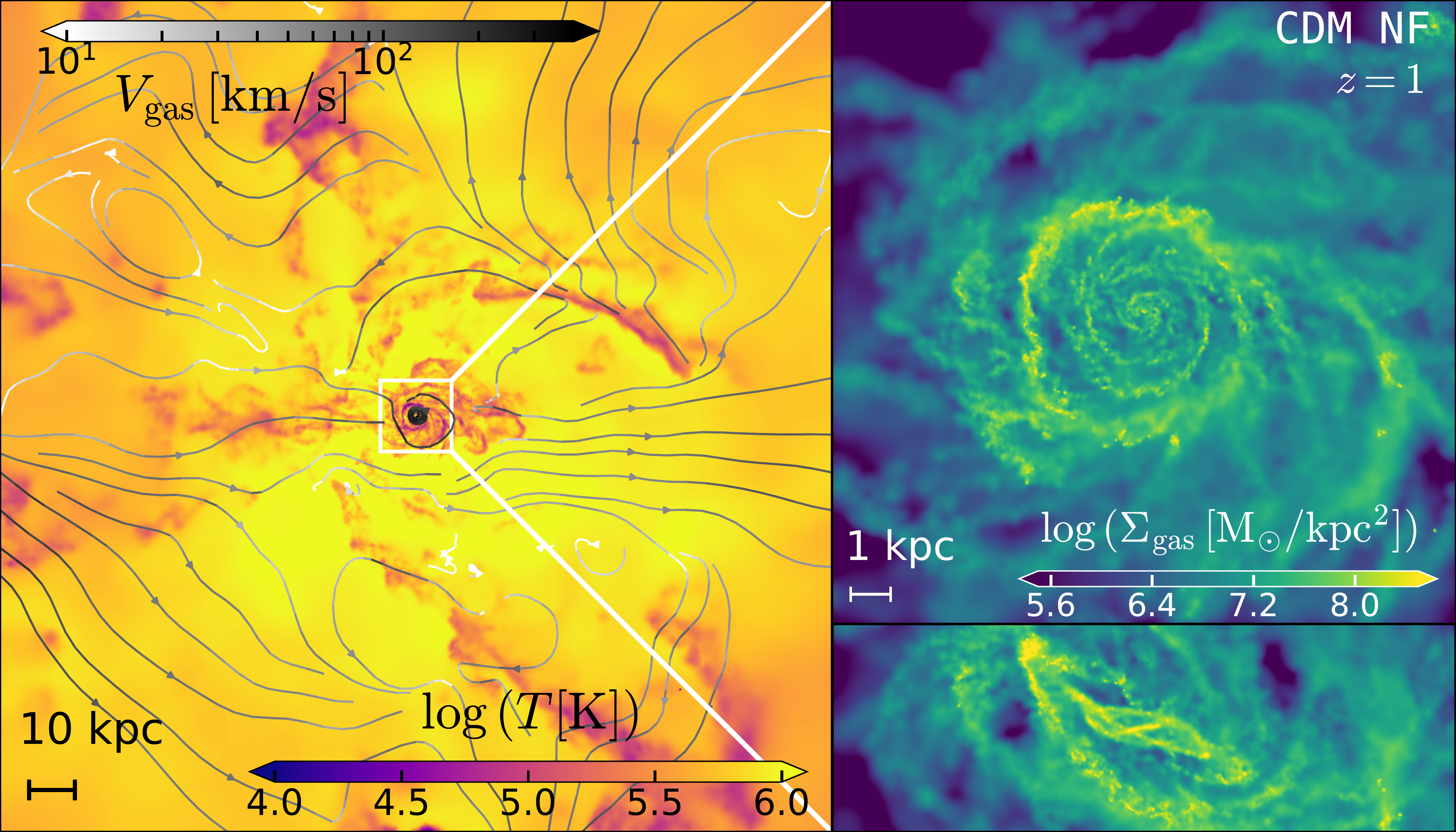}
    
    \caption{Four panels displaying the temperature and surface density distributions of ADM gas in $\admone$ and $\admtwo$ (left panels) and of baryonic gas in $\cdm$ and $\cdmnf$ (right panels). {\it Left of each panel:} Gas temperature distribution on the CGM scale at $z\sim 1$ ($350$ comoving $\kpc$ field-of-view). It is overlaid with the velocity field of the gas, where darker streamlines indicate higher velocities. Cooling physics and gravity determine the thermal properties of gas at this scale. Due to an order-of-magnitude stronger cooling efficiency, the ADM gas in $\admtwo$ is distinct from the other three cases, exhibiting stronger cooling flows at $z\sim 1$ and delayed inner CGM virialization. {\it Right of each panel:} Surface density of cold,  neutral gas in the central halo ($30$ comoving $\kpc$ field-of-view). Both face-on and edge-on views are shown. The neutral gas distributions are sensitive to gas accretion from the CGM, feedback from star formation (relevant for baryonic gas in the \texttt{CDM}, \texttt{ADM-1} and \texttt{ADM-2} simulations), as well as the thermal instability of the gas. For baryons in $\cdm$, an extended, co-rotating gaseous disk already forms at $z\sim 1$. For $\admone$ and $\admtwo$, we find compact ADM gas disks surrounded by streams of cold gas accreted from the CGM with poor alignment to the angular momentum of the central disk. The baryons in $\cdmnf$  exhibit a more extended gas disk than the ADM cases at this redshift, likely due to enhanced accretion via the helium peak in Fig.~\ref{fig:cooling_curves}.  Figure~\ref{fig:baryon_gas_mosaic} provides the corresponding panels for  $z\sim2$ and $z\sim 0$. Videos of the simulations are available \href{https://rb.gy/et2q0}{here}.} 
    \label{fig:baryon_gas}
\end{figure*}

This section describes the resultant distributions of ADM gas and clumps, as well as their effects on baryons and CDM. Throughout, we use several metrics to quantify the spatial morphology and kinematics of the distributions (Tab.~\ref{tab:Morphology_data}). When calculating these metrics, the coordinate system is aligned with the angular momentum of the youngest 25\% of baryonic stars within $\sim5\kpc$ of the galactic center.  

To characterize the kinematics, we use the distribution of orbital circularities (Fig.~\ref{fig:orbital_circ}), $\epsilon = j_z/j_c(E)$, defined as the ratio of the $z$-component of angular momentum to the angular momentum of a particle with the same energy on a circular orbit~\citep{Abadi2002}. Thin-disk orbits correspond to $\epsilon \geq 0.8$, thick-disk orbits to $0.2 \leq \epsilon < 0.8$, and spheroid-like orbits to $\epsilon < 0.2$~\citep{Yu2021}. The mass fraction of particles within the cylindrical region $|z|\leq 1.5\kpc$ and $r\leq 10\kpc$ in each of these $\epsilon$ ranges is denoted by $f_{\rm thin}$, $f_{\rm thick}$, $f_{\rm spheroid}$, respectively.

To characterize the spatial morphology, we define $z_i$ as the vertical height in which the $i^{\rm th}$ fraction of particles is contained in a cylindrical region of $r\leq 2 \kpc$.  Similarly, $r_{1/2}$ is the half-mass containment radius for particles within a cylindrical region $|z|\leq 2 z_{9/10}$.

\subsection{ADM Gas}
\label{sec:morphology}

In the standard picture of galaxy formation, gas collapses within the potential well of a DM halo, ultimately forming a rotationally-supported disk near its center.  Gas that accretes onto this central disk may have been previously shocked to the virial temperature or may have originated from cold, filamentary flows.  The stability of the resulting disk will be compromised if its self-gravity overwhelms its internal pressure. Disk  fragmentation is suppressed as the pressure support in the disk increases with the cut-off temperature of the cooling curve, which depends on the (dark) hydrogen binding energy.  

Figure~\ref{fig:baryon_gas} displays the $z \sim 1$ temperature and velocity distribution of the outer galactic gas (circumgalactic medium, CGM), on the left of each panel and the zoomed-in density profile of the central cold, neutral gas (interstellar medium, ISM), on the right of each panel. All species are cooling efficiently and forming high-density disks at this redshift. For the baryonic gas in both the $\cdm$ and $\cdmnf$ simulations, most of the CGM is shock-heated and at the virial temperature $\sim10^6 \K$; the same is true for the ADM gas in $\admone$. In contrast, the ADM gas in $\admtwo$  cools too rapidly to be held at the virial temperature and its CGM is dominated by high-density flows that free-fall into the ISM at $V_{\rm gas} \sim 400 \kms$, more than twice that of the other gas species. This is because the gas infall speed is approximately proportional to the cooling rate, which is greater for ADM-2.

The presence or lack of feedback also strongly influences disk formation.  As demonstrated by the surface density subpanels, the ADM gas in $\admone$ and $\admtwo$ accretes onto the central dark disk via dense filaments with poor alignment to its angular momentum. This is only possible because there is no feedback in the ADM sector to heat and disperse the filaments into the CGM. Moreover, the ADM disks are quite compact at this redshift  and are much smaller than the baryonic gas disk in $\cdm$; they are more comparable in size to the baryonic disk in $\cdmnf$ at $z\sim 1$, although the latter is still larger, possibly due to enhanced accretion from helium cooling.

We quantify gas fragmentation in the central galaxy ($r < 10\kpc$) at $z \sim 1$ by 
defining $\fgas$ as the fraction of total baryonic or ADM mass within 10~kpc that is made up of uncollapsed gas of that species. 
In $\cdm$ ($\fgas \approx 15\%$), the  baryons are significantly more gas-rich than those in $\cdmnf$ ($\fgas \approx 3\%$). The lack of feedback contributes to gas depletion because there is no mechanism to prevent runaway gravitational collapse. Similar depletion is observed for the ADM gas in $\admone$ ($\fgas \approx 4\%$) and $\admtwo$ ($\fgas \approx 6\%$).  
ADM-2 has a binding energy $\sim 3.5$ times greater than ADM-1, which may explain the relatively higher gas fraction.

\begin{figure*}
    \centering 
    \includegraphics[trim={0cm 0cm 0cm 0cm},clip,width=1.0\textwidth]{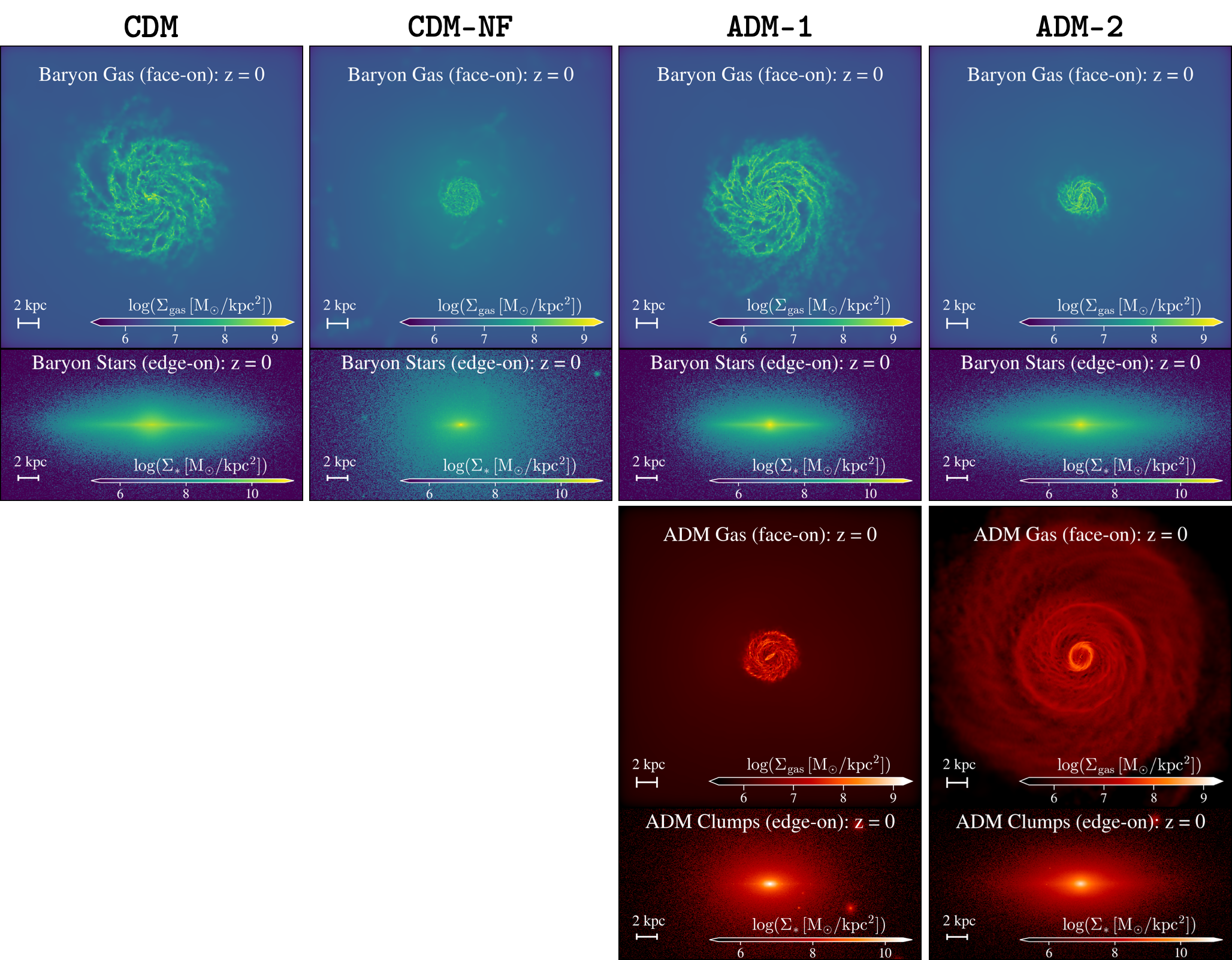}    \caption{\label{fig:final_baryon_images} Baryon and ADM density projection plots at the present day. The different simulations are organised by column, while the types of species are organised by rows. Gas is presented face-on while stars and ADM clumps are presented edge-on. The ADM gas disk in $\admone$ is as compact as the baryonic disk in $\cdmnf$ because both gas species have similar cooling rates, binding energies, and lack feedback. The ADM gas disk in $\admtwo$ undergoes enhanced accretion and is also more resistant to fragmentation, making it less compact. In both $\admone$ and $\admtwo$, $\sim95\%$ of the inner ADM is in the form of clumps, which are concentrated in the central-most regions of the galaxy.  Because of the lack of feedback, the ADM clumps in $\admone$ and $\admtwo$ and the baryonic stars in $\cdmnf$ lack a prominent thin disk and are more centrally concentrated. The large central densities of ADM in $\admone$ and $\admtwo$ enhance the central baryonic densities as well and in the case of $\admtwo$, the baryonic gas disk collapses catastrophically. Figures~\ref{fig:intermediate_morphologies}--\ref{fig:final_morphologies} provide the density distributions for additional redshifts, as well as for CDM. Videos of the simulations are available \href{https://rb.gy/et2q0}{here}.}
\end{figure*}

Figure~\ref{fig:final_baryon_images} presents the final $z = 0$ density projections for the baryonic and ADM species in all simulations. The ADM gas in $\admone$ forms a disk that is morphologically similar to that of baryons in $\cdmnf$. This is to be expected given the similar cooling timescales, identical binding energies, and lack of feedback. The ADM gas forms a more extended disk in $\admtwo$ than in $\admone$ due to increased cooling and central halo accretion rates, as well as lower fragmentation rates. 

In both $\admone$ and $\admtwo$, the ADM gas disks rotate along circular, gravitationally supported orbits. More than 90\% of the ADM gas 
 in each case is in a thin-disk  configuration with $\epsilon \geq 0.8$. Additionally, the angular momentum of the ADM gas disk is aligned with that of the baryonic gas disk in each simulation. The alignment is likely the result of two effects. Firstly, the initial density and velocity power spectra for baryons and ADM align closely with those of CDM, so the initial net angular momenta for ADM and baryons are similar. Secondly, the baryonic and ADM disks also likely maintain alignment through mutual gravitational torques, which dominate in the inner galaxy~\citep{Cadiou2021}.

\subsection{ADM Clumps}
\label{sec:sink_dynamics}

\begin{figure*}
    \centering 
    \begin{minipage}{0.9\textwidth}
    \centering
    \includegraphics[trim={1cm 1.75cm 1cm 0cm},clip,width=1.0\textwidth]{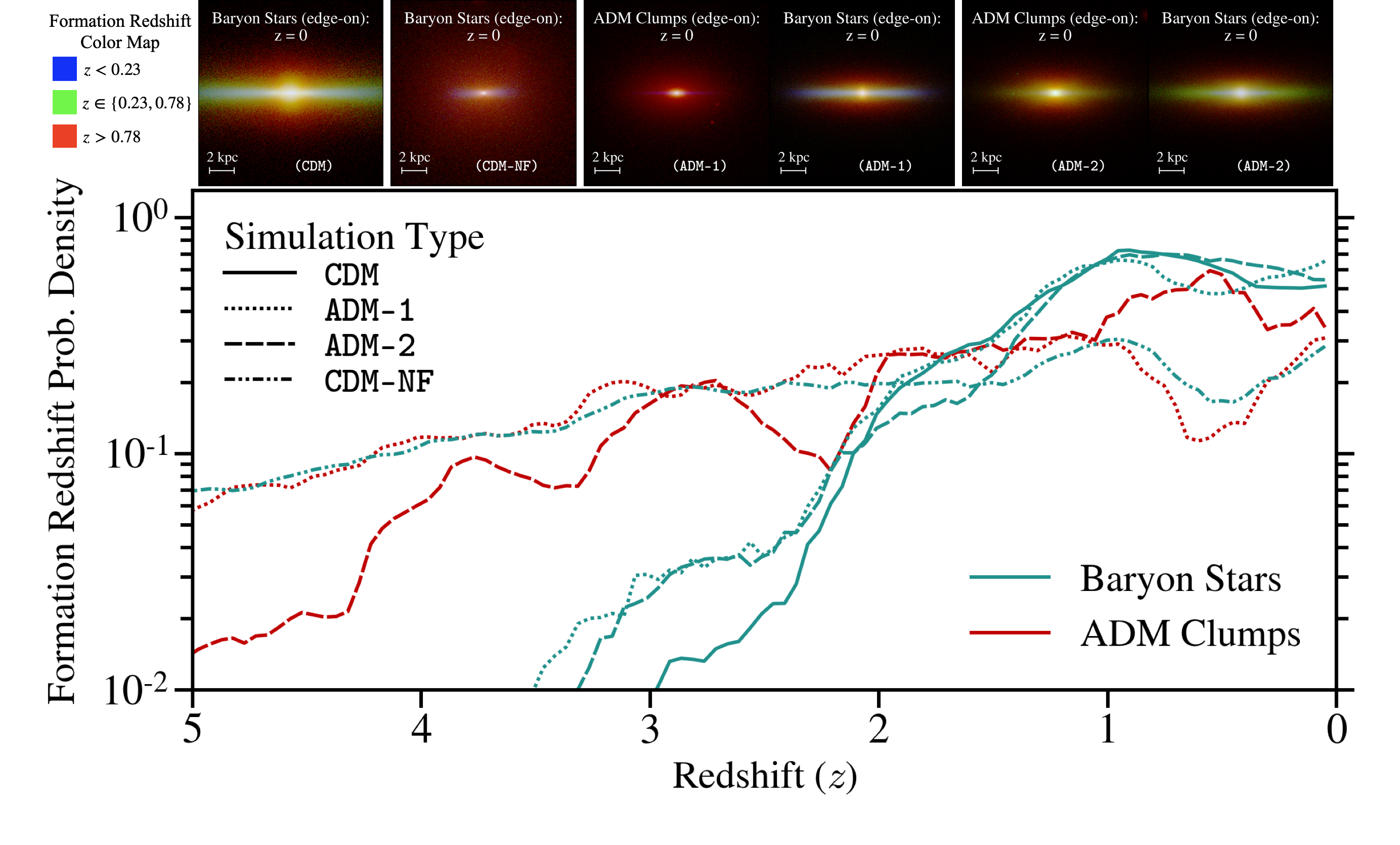}
    \end{minipage}
    \begin{minipage}{0.9\textwidth}
       \centering 
    \vspace{0.5cm}
    \includegraphics[trim={2cm 1cm 3cm 2.25cm},clip,width=1.0\textwidth]{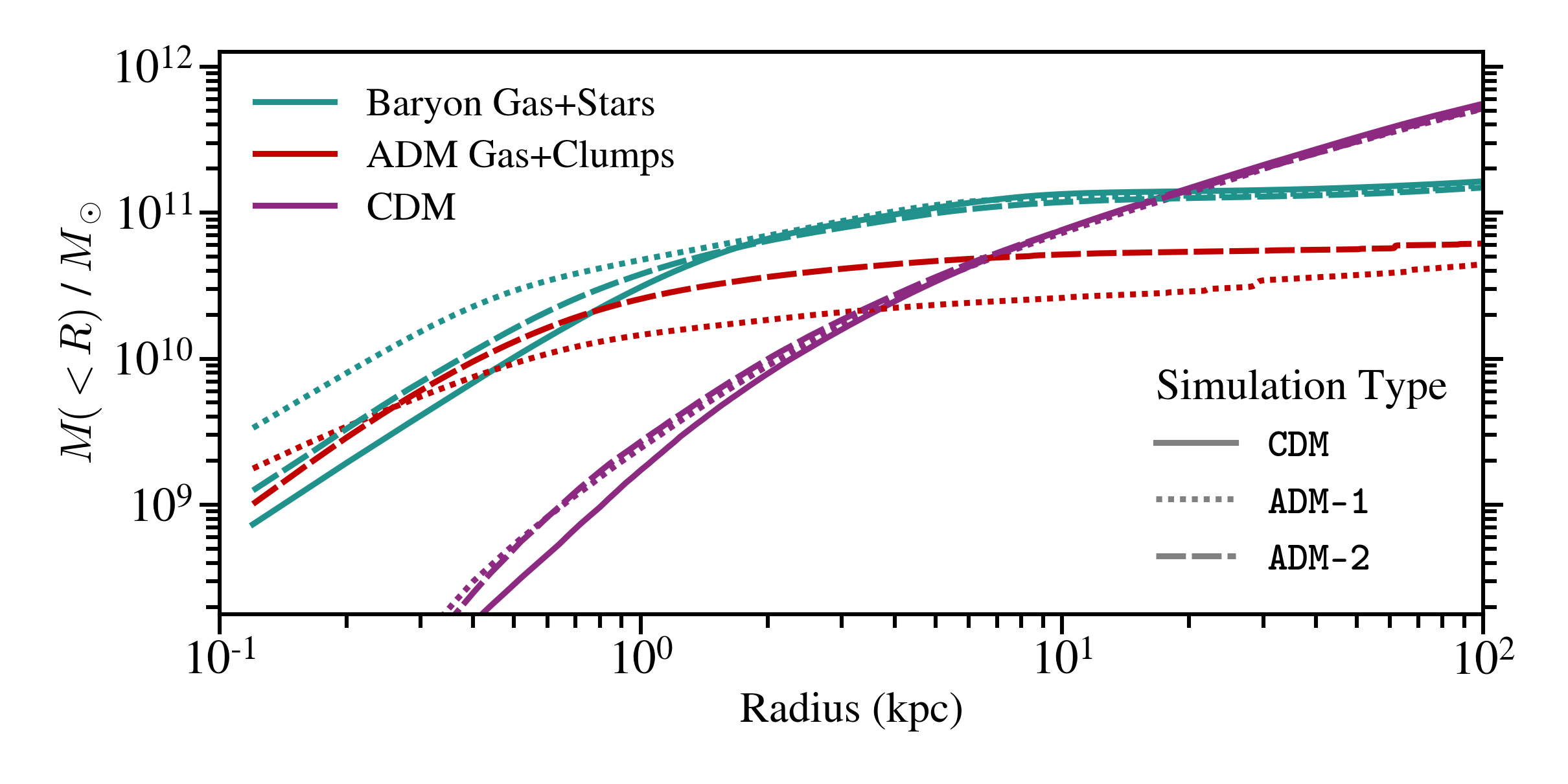}
    \end{minipage}
    \caption{\label{fig:sink_formation_ages} \emph{Top panel:} Formation history for baryonic stars and ADM clumps in the inner $10\kpc$ for all simulations. Colors represent the particle species and line styles represent the simulation. Above the figure are red-green-blue~(RGB) synthetic images for the particles of different ages (see the legend). The baryonic stars in $\cdmnf$ and the ADM clumps in $\admone$ and $\admtwo$ have greater formation probabilities at $z\gtrsim2$, compared to baryonic stars in $\cdm$.  This is also evident in the corresponding color maps which are red-dominant compared to that of the stars in $\cdm$.  \emph{Bottom panel:}~Enclosed mass profiles for all species at $z\sim0$. The solid, dotted and dashed lines reference the $\cdm, \admone$ and $\admtwo$ simulations, respectively. The different colors represent the separate particle species.  The ADM is concentrated within the inner $\sim\kpc$ of the galaxy, with an enclosed mass comparable to that of baryons.  The presence of this additional mass causes an overall enhancement in the baryonic and CDM mass in the central galaxy.}
    \label{fig:sfh_mass} 
\end{figure*}

When an ADM gas particle becomes so dense that it is Jeans-unstable, it converts into a clump.  This is similar to the process by which baryonic gas gets converted into stars, except that fewer formation criteria are applied in the ADM case. ADM clump particles have masses of $1.67\times 10^4~\mstar$, while baryonic star particles have masses of $5.6\times 10^4~\mstar$.  Both are collisionless and do not interact hydrodynamically with the remaining gas. At $z\sim0$, stars dominate the central density of the galaxy, accounting for 95\% of the total baryonic mass within the inner $10\kpc$; similar fractions of ADM are in clumps versus gas in this region.

Figure~\ref{fig:sink_formation_ages} illustrates the formation history of the baryonic stars and ADM clumps in the inner $10\kpc$ for all simulations. The figure also includes images showcasing the edge-on projections of the baryonic stars and ADM clumps at $z\sim0$, synthesized with surface density maps of stars/clumps in three age bins. 

In $\cdm$, 60\% of the baryonic stars within the inner $10\kpc$ form at $z \lesssim 1$ when the galaxy's evolution is relatively quiet~\citep{daniel2016,Stern2020,Hafen2022,Yu2021}. Star formation is suppressed at earlier redshifts because the gravitational potential of the halo is not deep enough to prevent dense gas from being ejected by supernovae feedback~\citep{Hopkins2023}.  The corresponding color-map image shows that baryonic stars separate out visually: thin-disk stars are blue, thick-disk stars are green, bulge stars are a mix of blue and green (appearing yellow), and halo stars are red. In $\cdmnf$, baryonic star formation occurs more uniformly across redshifts and only 23\% of stars form at $z\lesssim 1$.
The fraction of stars in the thick disk is enhanced relative to that in $\cdm$ ($\sim53\%$ versus  $39\%$). The older distribution of stars is also evidenced by the redder edge-on color map. Stars formed during the early, bursty periods of star formation in FIRE-2 simulations have been shown to dominate the thick disk~\citep{Yu2021}. 

The spatial and kinematic distributions of ADM clumps in $\admone$ and $\admtwo$ resemble those of baryonic stars in $\cdmnf$. In both simulations, the ADM clumps start to form at $z\gg 1$, and their evolution is characterized by periods of ``bursty'' formation following merger events. The impact of bursty formation events is even more prominent since the ADM gas accounts for a lower total mass fraction than baryonic gas and thus has a lower average density. Because the ADM clumps form early, as evidenced by their deep red color maps, they are  kinematically hotter, with a larger proportion having thick-disk and spheroidal circularities. For both $\admone$ and $\admtwo$, $\lesssim 20\%$ of the clumps have orbital circularities consistent with a thin disk, compared to $\sim 38\%$ for baryonic stars in $\cdm$.  

Generally, the ADM clumps are concentrated within the inner $\sim\kpc$ of the galaxy, as demonstrated by the enclosed mass profiles in Fig.~\ref{fig:sfh_mass}. 
 In this central region, the total enclosed mass of ADM is almost an order-of-magnitude greater than that of CDM.    The spatial distributions of ADM-1 and ADM-2 clumps are also more concentrated, with $r_{1/2}$ values nearly 2--3 times smaller than that of baryonic stars in $\cdm$. This is reasonable since, in the absence of feedback, ADM clumps and stars can form in greater quantities in the central galaxy where densities are higher. The $\rhalf$ for baryonic stars in $\cdmnf$ is greater than that of ADM clumps in $\admone$ and $\admtwo$. One explanation is that baryonic cooling is enhanced due to helium,
 while ADM cooling is artificially suppressed without dark molecular cooling, 
allowing the baryonic gas disk to cool and collapse into stars at larger radii.

\subsection{ADM Backreaction on Baryons and CDM}
\label{sec:baryons}

The ability of ADM gas to cool efficiently without feedback to limit collapse means that it dominates the inner galactic densities at early times, which has important consequences for baryons.  At $z \sim 4$ in $\admone$ and $\admtwo$, the mass of ADM gas in the inner $\sim0.5\kpc$ of the galaxy is already more than ten times greater than that of baryons~(Fig.~\ref{fig:central_density_evolution}). 

The build-up of ADM mass deepens the central gravitational potential, leading to more centrally concentrated baryonic distributions, as is visually apparent in Fig.~\ref{fig:final_baryon_images} and Figs.~\ref{fig:density_profiles}--\ref{fig:adm_density_profiles}. 
Compared to $\cdm$, at $z \sim 0$ the total enclosed mass of baryons within $\lesssim 1\kpc$ in $\admone$~($\admtwo$) is enhanced by $\gtrsim 50\%$~($\gtrsim 16\%$) in Fig.~\ref{fig:sfh_mass}.  Beyond this radius, the baryonic mass profiles in these three simulations are essentially indistinguishable. In general, the distributions of baryonic gas and stars in $\admone$
have lower $\rhalf$ by $\gtrsim 10$\%. However, the effect is more dramatic for $\admtwo$: the baryonic gas disk collapses and depletes to $\sim 30\%$ 
 of its $\rhalf$ value in $\cdm$, almost certainly excluding this particular ADM scenario.

The CDM distribution is also contracted, with the enclosed mass within $\sim1\kpc$ enhanced by $\gtrsim 40\%$ for both $\admone$ and $\admtwo$. The total enhancement to the inner enclosed mass can have observable effects on the rotation curve and the galactic bar. The greater inner galactic rotation speeds (Fig.~\ref{fig:circular_v_profiles}) may place both ADM scenarios in tension with galactic velocity measurements~\citep{Sohn2017} although higher resolution simulations are necessary to confirm this.

Lastly, the enhancement of the central baryonic density due to an efficiently cooling, dense ADM core, noticeably enhances star formation at early times:
the fraction of stars that form at large redshift $z\in\{2,10\}$ in $\cdm$, $\admone$ and $\admtwo$ is $\approx$ 4\%, 7\% and 6\%, respectively. 

\section{Conclusions}

This Letter presents the first cosmological hydrodynamical simulations with multiple DM species: CDM as well as a dissipative subcomponent, ADM.  We developed an ADM module for~\texttt{GIZMO} and simulated Milky Way-analogues.  
We considered two benchmark ADM scenarios with distinctive cooling properties that affect the accretion rate onto the central galaxy as well as gas fragmentation.  In both cases, the ADM has no self-driven feedback. 

For each benchmark, the ADM gas cools efficiently and forms a central gas disk aligned with the baryonic stellar disk in the galaxy.  Only a small fraction of the ADM in the central galaxy exists in cold, neutral gas.  The vast majority ($\gtrsim 95\%$) is in collisionless clumps of collapsed gas.  These ADM clumps form relatively early in the galaxy's evolution and sink to its center, where their total mass in the inner-most $\kpc$ of the galaxy is comparable to that in baryons.  
The collapse of ADM at early times deepens the central galactic potential, in turn significantly increasing the central density of  baryons and CDM. This affects the rotational kinematics of the inner galaxy and increases baryonic star formation at earlier redshifts ($z \gtrsim 2$). In \admtwo, the baryonic gas disk is severely depleted compared to the observed Milky Way, very likely excluding this scenario. In \admone, the enhanced inner rotation curve could place it in tension with galactic velocity measurements~\citep{Sohn2017}, but higher resolution simulations are necessary to confirm this.

The distribution of ADM gas near the Solar position is relevant for direct detection searches~\citep{Chacko:2021vin, Curtin:2020tkm} and can be lower than the naive expectation set by the ADM fraction.  This is because the majority of ADM gas that collapses into a disk ultimately fragments into centrally-concentrated clumps. 
%
Gravitational signatures of the dark disk depend sensitively on its morphology. For ADM-1 and ADM-2, the
thin
dark-disk constraints from~\citet{Kramer:2016dqu, Schutz2017, Buch:2018qdr,Widmark2021} do not apply, but microlensing searches will be highly sensitive~\citep{Winch:2020cju}.

%
%


%
Our simulations will enable quantitative predictions for specific galactic observables that may reveal the presence of an ADM subcomponent.  These include: modifications to the baryonic stellar disk, differences in stellar formation histories, changes to the subhalo mass function, and novel properties of dwarf galaxies.  When extended to include molecular cooling, such simulations will also lead to an improved understanding of the populations of dark compact objects that result from dissipative interactions.

The study of strongly dissipative DM has long been hampered by the challenges of making reliable predictions on galactic scales. This work makes possible a wide variety of new investigations that will illuminate the astrophysical dynamics of this generic class of DM.

\section{acknowledgments}
This work would not be possible without the helpful comments and expertise of the following individuals: Tom Abel, Jared Barron, Peter Berggren, Malte Bushmann, Francis-Yan Cyr-Racine, Ben Dodge, Kareem El-Badry, Caleb Gemmell, Akshay Ghalsasi, James Gurian, Donhui Jeong, Laura Keating, Andrey Kravtsov, Hongwan Liu, Lina Necib, Michael Ryan, Sarah Shandera, Oren Slone, David Spergel,  Romain Teyssier, Francisco Villaescusa-Navarro and Andrew Wetzel. A special thanks goes to Jack Setford for theoretical discussions and support during the early stages of the project. We are also grateful to JiJi Fan and Lisa Randall for their helpful comments on our draft. ML and SR are supported by the Department of Energy (DOE) under Award Number DE-SC0007968. ML also acknowledges support from the Simons Investigator in Physics Award. 
The research of DC was supported in part by a Discovery Grant from the Natural Sciences and Engineering Research Council of Canada, the Canada Research Chair program, the Alfred P. Sloan Foundation, the Ontario Early Researcher Award, and the University of Toronto McLean Award.
NM acknowledges the support of the Natural Sciences and Engineering Research Council of Canada (NSERC) funding reference number RGPIN-2017-06459. This work was performed in part at the Aspen Center for Physics, which is supported by the National Science Foundation (NSF) grant PHY-1607611. Support for XS \&\ PFH was provided by the NSF Research Grants 1911233, 20009234, 2108318, the NSF Faculty Early Career Development Program (CAREER) grant 1455342, the National Aeronautics and Space Administration (NASA) grants 80NSSC18K0562, HST-AR-15800. 

Numerical simulations were run on the supercomputer Frontera at the Texas Advanced Computing Center (TACC) under the allocations AST21010 and AST20016 supported by the NSF and TACC, and NASA HEC SMD-16-7592. Early testing and analyses were done on the Niagara cluster.  The work presented in this paper was performed on computational resources managed and supported by Princeton Research Computing. This research made extensive use of the publicly available codes
\texttt{IPython}~\citep{PER-GRA:2007}, 
\texttt{Jupyter}~\citep{Kluyver2016jupyter}, \texttt{matplotlib}~\citep{Hunter:2007}, 
\texttt{NumPy}~\citep{harris2020array}, 
\texttt{SciPy}~\citep{2020SciPy-NMeth}, \texttt{SWIFTsimIO}~\citep{Borrow2020}, \texttt{unyt}~\citep{Goldbaum2018}, 
\texttt{gizmo-analysis}~\citep{Wetzel2020} and \texttt{Python Imaging Library}~\citep{clark2015pillow}.

\bibliography{simulating_adm}

\newpage

\appendix

\section{ADM Cooling Rates}
\label{sec:dark_cooling_rates}

\setcounter{equation}{0}
\setcounter{figure}{0} 
\setcounter{table}{0}
\renewcommand{\theequation}{A\arabic{equation}}
\renewcommand{\thefigure}{A\arabic{figure}}
\renewcommand{\thetable}{A\arabic{table}}

The Atomic Dark Matter~(ADM) model considered in this work consists of a dark electron, $e'$, dark proton, $p'$, and a massless dark photon, $\gamma'$. In this paper, we consider the scenario where the majority of the dark matter~(DM) exists in Cold Dark Matter~(CDM), and a subdominant fraction ($6\%$) is in ADM.  While the CDM forms stable halos, the ADM acts in wholly novel ways given its strongly dissipative nature.  In the early Universe, the dark electrons and protons combine to form neutral, bound-state dark hydrogen, $H_0'$, with binding energy $B_0^{'} = \frac{1}{2}{\alpha'}^2 m_{e'} c^2$, within a galaxy's host halo.  As the galaxy evolves, however, the ADM gas is shock-heated to the virial temperature and ionized.  Therefore, the ADM in a galaxy exists as a mixture of both neutral gas and  ionized plasma.  The fraction in each is set by the cooling rate in the dark sector, which depends on the fundamental parameters of the ADM model.    

We use the atomic cooling equations derived in~\citet{Rosenberg2017}~(see also, \citet{Ryan:2021dis,Ryan:2021tgw,Gurian:2021qhk}) and neglect molecular cooling.  The latter assumption suppresses gas collapse rates in the central galaxy at late times since gas temperatures cannot decrease below the dark hydrogen binding energy. The inclusion of  molecular cooling processes is necessary for detailed modeling of the dark interstellar medium, which is beyond the scope of this work. In more detail, the ADM cooling processes considered here include the following: 
\begin{itemize}
    \item Collisional excitation: 
    $H_0'\, + \, e' \rightarrow H_0'^*\, + \, e' \rightarrow H_0'\, + \, e'\, + \,  \gamma'$\, 
    \item Collisional ionization: 
    $H_0'\, + \, e' \rightarrow e'\, + \, p'\, + \, e'$
    \item Inverse Compton Scattering (from dark CMB): 
    $e' \, +\, \gamma' \rightarrow e'\, + \, \gamma'$
    \item Recombination: 
    $e'\, + \, p' \rightarrow H_0'\, + \, \gamma'$
    \item Bremsstrahlung: 
    $e' \, +\, p' \rightarrow e' \, +\, p'\, + \, \gamma'$ ,
\end{itemize}
where $H_0'^*$ is an excited state.  Collisional excitation of dark hydrogen dominates the cooling rate at low temperatures, while Bremsstrahlung dominates at high temperatures, 
\begin{equation}
    T \ \gtrsim  \ \left(\frac{{\alpha'}^2 m_{e'}}{\alpha^2 m_e}\right) \  \times  \ (10^6\K ) \, ,
\end{equation}
where the ADM is mostly ionized (see \citet{Ryan:2021tgw} for baryonic cooling rate re-scalings).

For process $i$ with velocity-dependent cross section, $\sigma_i(v)$, the cooling rate is given by 
\begin{equation}
\Gamma_i(T) = \langle\sigma_i \, v\rangle \, n_A n_B \, ,
\label{eq:rate}
\end{equation}
where $n_{A,B}$ are the number densities of the initial species, assumed to move with some relative speed $v$.  The corresponding cooling rate is
\begin{equation}
    P_{i,j}(T) = \langle E_j \, \sigma_i \, v\rangle \, n_A n_B \, ,
    \label{eq:power}
\end{equation}
where $E_j$ is the energy of the initial state in either species $j = A,B$.  The thermal averages in 
Eqs.~\ref{eq:rate}--\ref{eq:power} are performed assuming that the dark electron speeds at temperature $T$ are well-modeled by a Maxwell-Boltzmann distribution,  
\begin{equation}
    f(v, T) = 4 \pi v^2 \, \left(\frac{m_{e'}}{2 \pi T}\right)^{3/2} \, \exp\left[\frac{-m_{e'} \, v^2}{2T}\right] \, ,
\end{equation}
where $m_{e'}$ is the dark electron mass.  Because we only consider parameters where the dark proton is much more massive than the dark electron ($\mphat \gg \mehat$), the relative speed between the $e'$ and either $p'$ or $H'$ will be approximately equal to the dark electron speed.

\begin{figure}
    \centering 
    \includegraphics[trim={0cm 0cm 0cm 0cm},clip,width=0.49\textwidth]{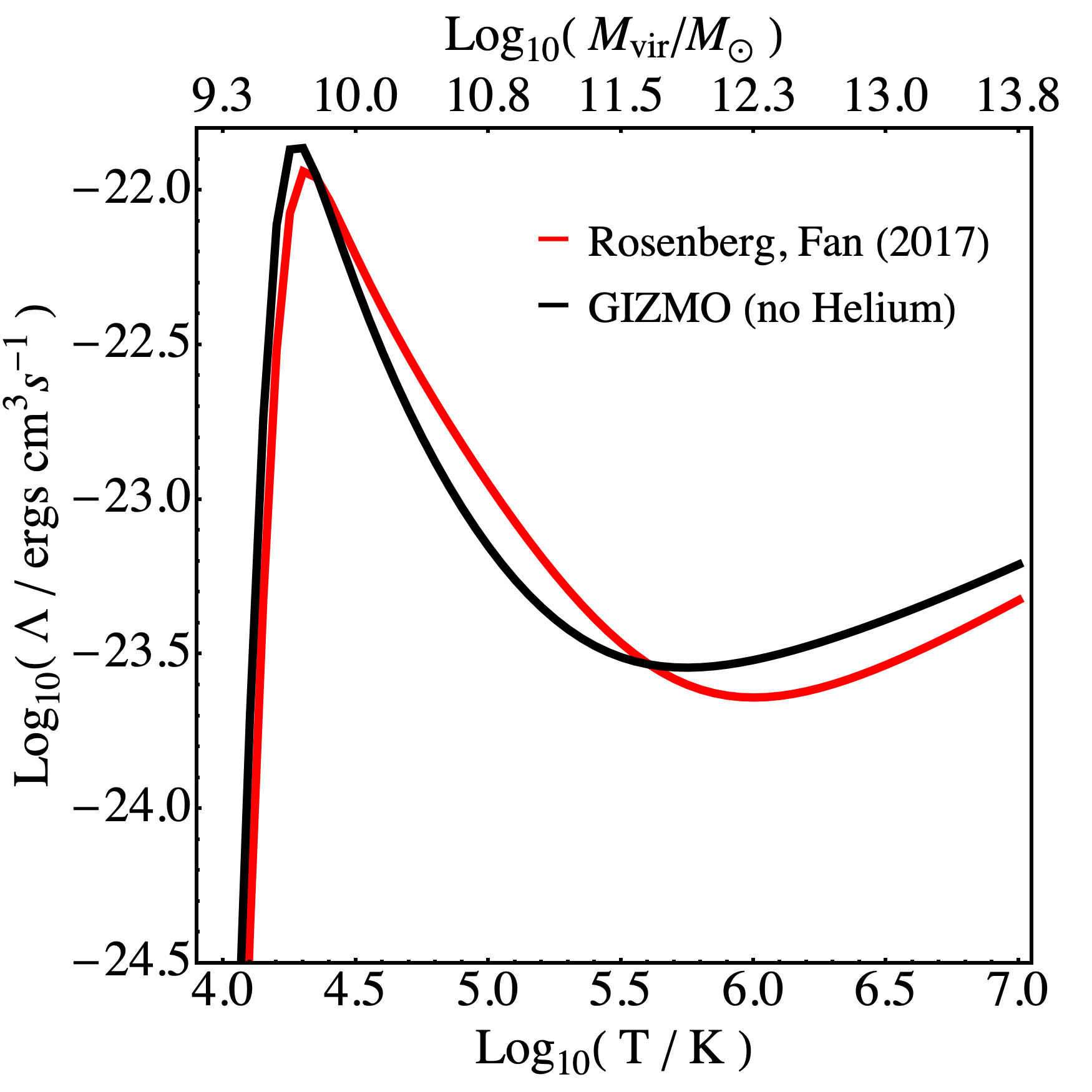}
    \includegraphics[trim={0cm 0cm 0cm 0cm},clip,width=0.49\textwidth]{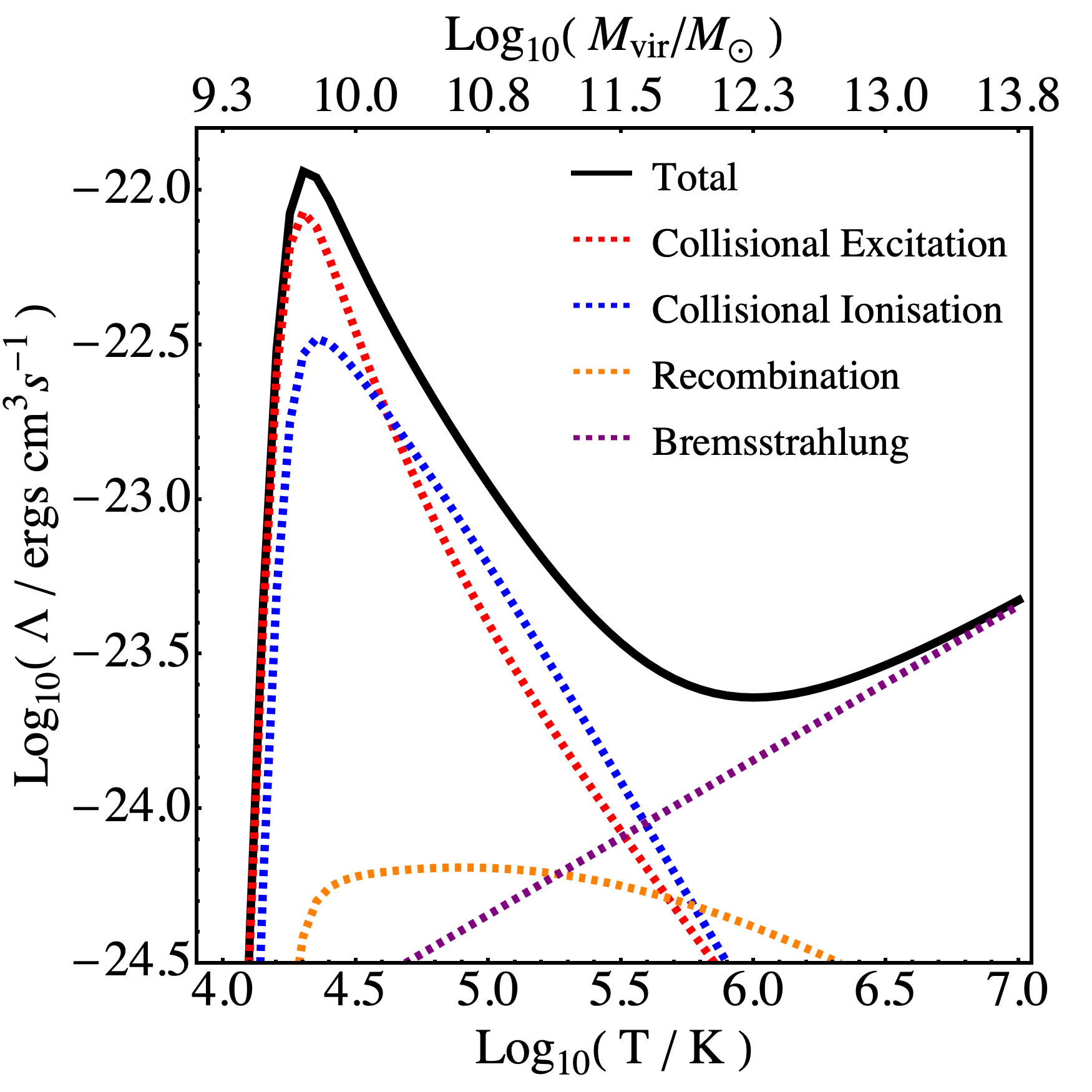}
    \caption{\label{fig:cooling_rate_verification} \emph{Left panel:} Comparison between the cooling rates used in \texttt{GIZMO} \citep{Hopkins2018}, excluding metals like helium, and those of \citet{Rosenberg2017} with $\mehat=m_{\rm e}$, $\mphat=m_{\rm p}$ and $\alphahat=\alpha$. Just as in Fig.~\ref{fig:cooling_curves}, the upper axis translates the temperature $T$ to the virial temperature of a galaxy with virial mass $M_{\rm vir}$ \citep{Bryan1997}. These curves agree to $\order{1}$, which is sufficient for the purposes of this investigation. \emph{Right panel:} Breakdown of the contribution of different cooling processes to the total \citet{Rosenberg2017} cooling rate in the left panel. Collisional excitation dominates at low temperatures $\log_{10}(T/\K) \in \{4.2,5.0\}$, while Bremsstrahlung (i.e. free-free cooling) dominates at high temperatures $T \gtrsim 10^6\K$.}
\end{figure}

\citet{Rosenberg2017} detail the assumptions that are made to obtain their ADM cooling rates, and we refer the interested reader to their work for a comprehensive overview. Here, we highlight three of the most important assumptions.  The first is that the 
ADM gas is optically thin, so that any dark photons emitted in the cooling process are not reabsorbed.  Second, is that $m_{e'} \ll m_{p'}$, which---as described above---has implications when modeling the relative speeds of the different species, but also means that the reduced mass of the dark hydrogen can be approximated as simply $m_{e'}$.  Lastly, the calculated rates omit the $\mathcal{O}(1)$ Gaunt factors, so one should consider the resulting rates to be the correct order-of-magnitude, but not exact. This approximate agreement is seen in Fig. \ref{fig:cooling_rate_verification} where we compare the standard cooling rates to those of \citet{Rosenberg2017} with the Standard Model particle parameters. 

To compute the complete cooling rates in Eq.~\ref{eq:power}, we need the number densities of the interacting species. To do this, we assume that the collisional ionisation~(CI) rate balances the  recombination~(R) rate: 
\begin{equation}
\langle \sigma_{\scriptscriptstyle{\rm CI}} v\rangle \, \xHneutralhat \xehat \sim \langle\sigma_{\scriptscriptstyle{\rm{R}}} v\rangle \,   \xehat \xphat \, ,    
\end{equation}
where $x_i = n_i/\nHhat$, with $\nHhat = \nHneutralhat + \nphat$, and the ADM gas is assumed to be neutral so that $\nehat = \nphat$ and thus $1=\xphat +\xHneutralhat= \xehat +\xHneutralhat$. It thus follows that 
\begin{equation}
    \label{eq:CIE_solutions}
    \xehat \sim \frac{\langle\sigma_{\scriptscriptstyle{\rm{CI}}} v\rangle}{\langle\sigma_{\scriptscriptstyle{\rm{CI}}}v\rangle+\langle\sigma_{\scriptscriptstyle{\rm{R}}} v\rangle} \quad\quad \text{and} \quad \quad 
    \xHneutralhat \sim  \frac{\langle\sigma_{\scriptscriptstyle{\rm{R}}} v\rangle}{\langle\sigma_{\scriptscriptstyle{\rm{CI}}} v\rangle+\langle\sigma_{\scriptscriptstyle{\rm{R}}} v\rangle} \, .
\end{equation}
This assumption, namely ionisation equilibrium (IE), is valid for the densities typical in Milky Way-mass galaxies. We verify the validity of this assumption in Sec.~\ref{sec:IE_equilibrium_validation}.

\begin{figure}
    \centering 
    \includegraphics[trim={0cm 0cm 0cm 0cm},clip,width=0.49\textwidth]{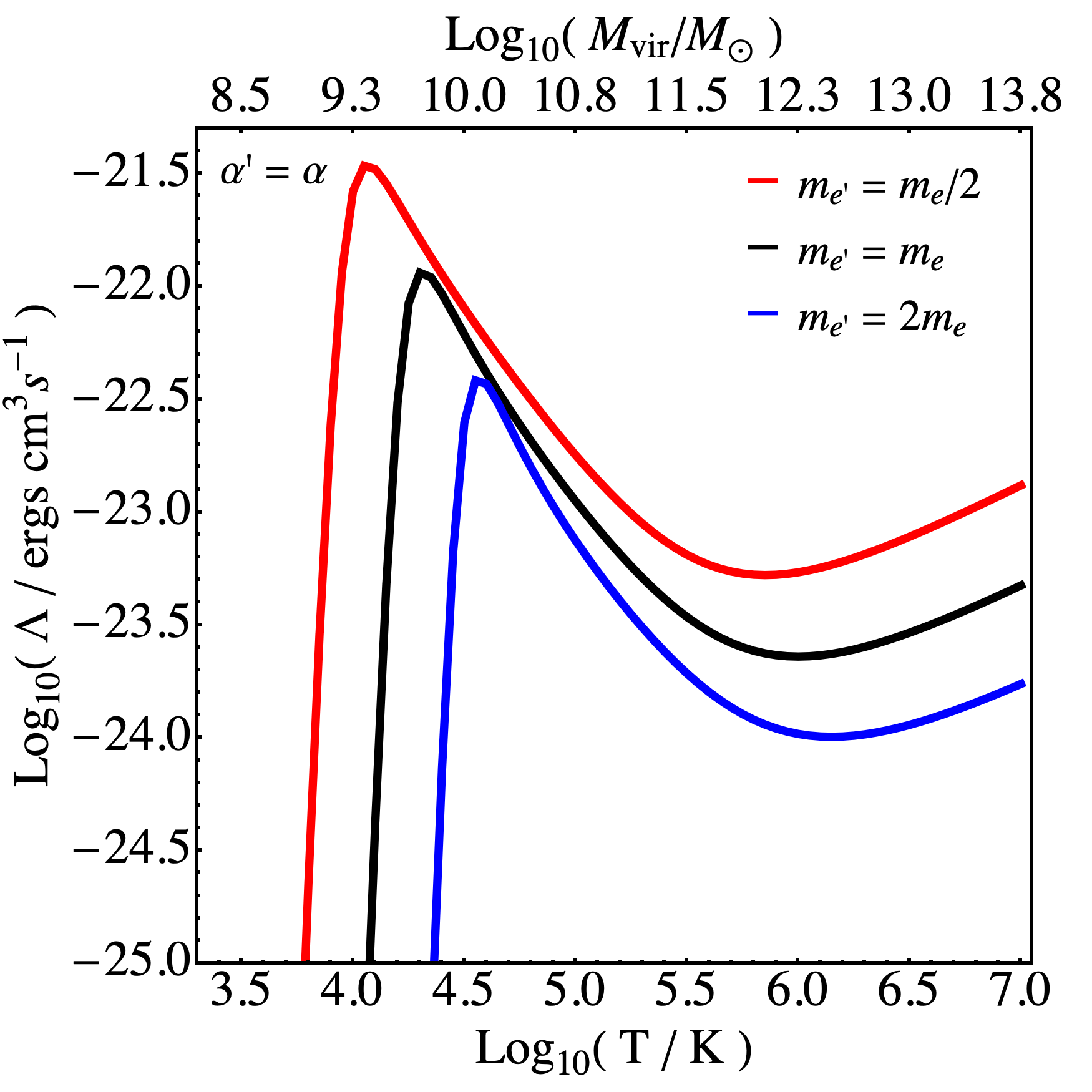}
    \includegraphics[trim={0cm 0cm 0cm 0cm},clip,width=0.49\textwidth]{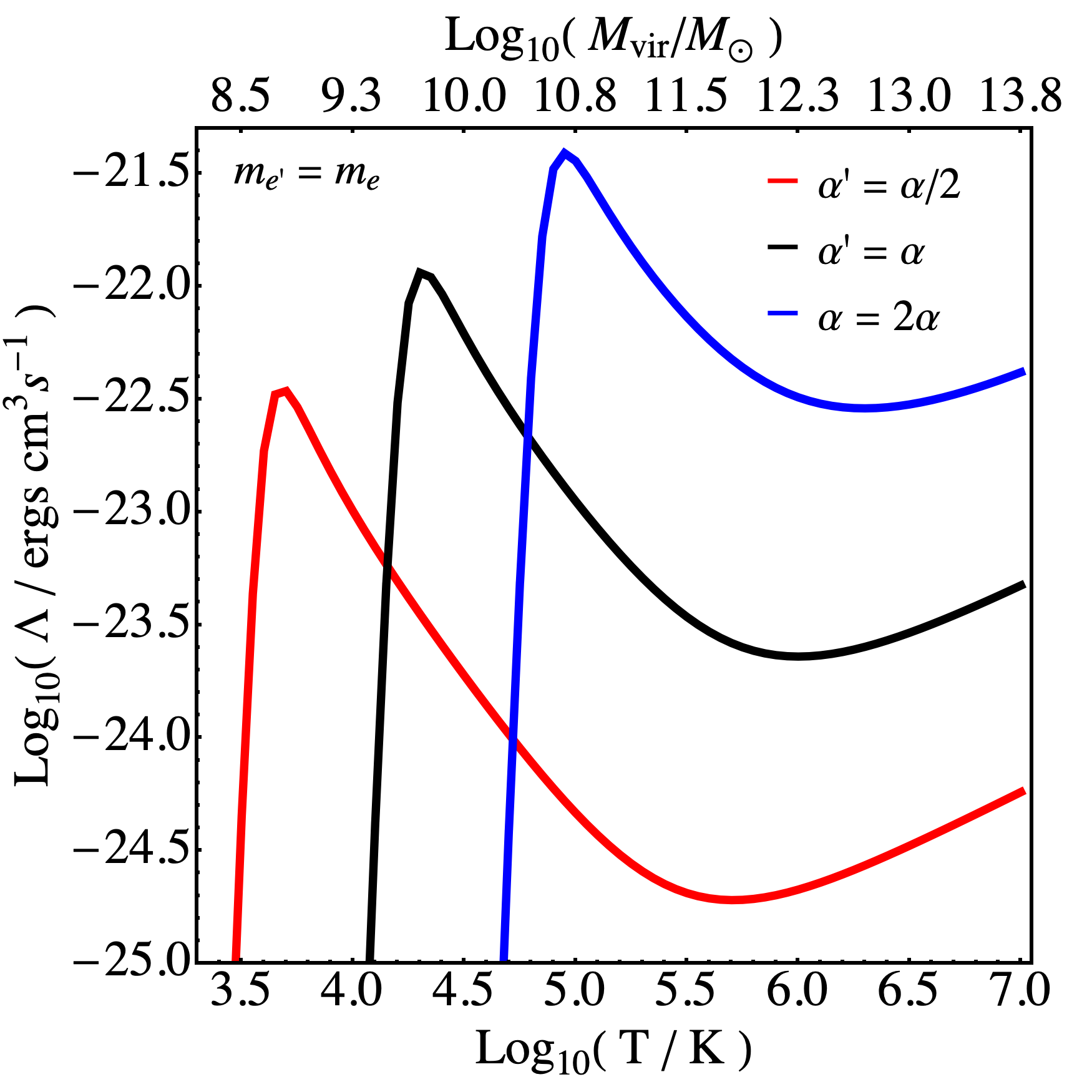}
    \caption{\label{fig:cooling_rate_varying_params} \emph{Left panel:} Cooling rates with varying $\mehat$ and fixed $\alpha' = \alpha$. Just as in Fig.~\ref{fig:cooling_curves} and Fig.~\ref{fig:cooling_rate_verification}, the upper axis translates the temperature $T$ to the virial temperature of a galaxy with virial mass $M_{\rm vir}$ \citep{Bryan1997}. As $\mehat$ is reduced, the cut-off temperature decreases, since it scales as the dark hydrogen binding energy, which is proportional to $\mehat$. The peak  cooling rates increase because less massive electrons interact more efficiently. \emph{Right panel:} Cooling rates with varying $\alphahat$ and fixed $\mehat = m_e$. As $\alphahat$ decreases, the cut-off temperature decreases because the dark hydrogen binding energy scales as $\alphahat^2$. The peak cooling rates also decrease for weaker coupling constants.}
\end{figure}

Figures \ref{fig:cooling_rate_verification} and \ref{fig:cooling_rate_varying_params} show the breakdown of ADM cooling processes and how the total rate varies with the choice of parameters. 
The right panel of Fig.~\ref{fig:cooling_rate_verification} shows  how the different cooling processes (collisional excitiation and ionization, inverse Compton scattering, recombination, and Brehmsstrahlung) contribute to the total cooling rate for ADM with $\mehat=\me$, $\mphat = \mproton$ and $\alphahat=\alpha$. Collisional excitation dominates at low temperatures $10^{4}\K \lesssim T \lesssim 10^{5}\K$, while Bremsstrahlung dominates at higher temperatures $T\gtrsim 10^6\K$. In Fig.~\ref{fig:cooling_rate_varying_params}, one can see that varying the $\mehat$ and $\alphahat$ alters the cut-off temperature since it scales with the binding energy, $\sim \mehat\alphahat^2$. The peak collisional excitation cooling rate and ionised free-free cooling rates also increase monotonically with decreasing $\mehat$ and increasing $\alphahat$ at temperatures above $\sim 0.1 B_0'$. 

\section{Validating Ionisation-Recombination Equilibrium}
\label{sec:IE_equilibrium_validation}

The reason that ionisation equilibrium holds for galactic densities and temperatures is that the typical ionisation and recombination timescales are much smaller than the dynamical timescales.  In particular, the IE timescale is approximately
\begin{equation}
    \tIE \sim \frac{1}{n_{e'} \, (\Gamma_{\scriptscriptstyle{\rm CI}} + \Gamma_{\scriptscriptstyle{\rm R}})} 
    \sim \frac{1}{\nHhat \Gamma_{\scriptscriptstyle{\rm CI}} } \, ,
\end{equation} 
where the last approximation follows from the  assumption of Eq.~\ref{eq:CIE_solutions}.
The other relevant timescale is the dynamical time (or free-fall time),
\begin{equation}
\tff \sim \sqrt{\frac{3\pi}{32 \, G \, \rho}} \sim \sqrt{\frac{3\pi}{32 \, G \, \nHhat \, \mphat}} \, ,  
\end{equation}
where $G$ is Newton's gravitational constant and $\rho$ is the average density of the dark gaseous halo. For IE to be valid, $\tIE \ll \tff$. 

We can check that this is valid for the densities and temperatures typical of the Milky Way-like galaxy \texttt{m12i}. For the densities and temperatures typical of the circumgalactic medium~(CGM) of the galaxy, fitting a simple power law to the baryonic gas  in $\cdm$ gives the following  fit for $r\in[10,100]\kpc$: 

\begin{align*}
    \rho_{\rm baryon}(r) \approx 10^{3.4} \msun \kpc^{-3} \, \left(\frac{r}{100\kpc}\right)^{-1.87} \quad \quad \mbox{ and} \quad \quad
    T_{\rm baryon}(r) \approx 10^{5.8} \K \, \left(\frac{r}{100\kpc}\right)^{-0.75} \, .
\end{align*}

For ADM, we use the same functional forms, but scaled as $\rho_{\rm adm} \approx \rho_{\rm baryon}\,\Omega_{\rm adm}/\Omega_{\rm b}$ and $T_{\rm adm} \approx T_{\rm baryon}\, \mphat/m_{\rm p}$ (since $T_{\rm vir} \propto G M_{\rm vir} \mphat/R_{\rm vir}$). We do not fit directly to the ADM gas in $\admone$ and $\admtwo$ since the simulation cooling rates explicitly rely on ionisation-recombination equilibrium; however, the approximate re-scalings presented here do provide a reasonable fit to the simulated ADM gas distributions. Calculating the ratio $\tIE/\tff$ for the ADM-1 and ADM-2 parameters gives us $\tIE/\tff \in [10^{-6},10^{-5}]$ between 10 and 100 kpc. This validates the use of ionisation-equilibrium in the ADM cooling module. Within the dark ISM at $r<10\kpc$, the temperatures will be near the peak of the cooling curve in Fig.~\ref{fig:cooling_curves} and will have much greater densities than those in the dark CGM. Thus, ionisation-equilibrium will also hold for the efficiently cooling gas in the dark ISM.


\section{Brief Overview of GIZMO Implementation of ADM Physics}
\label{sec:gizmooverview}

ADM is implemented in \texttt{GIZMO} as a separate gas species. This is achieved using a separate identification number in the code's gas structure: \texttt{ADMType}. It is an integer with different values for ADM and for baryons. Within GIZMO's hydrodynamic routine, we separate ADM and baryons by requiring that all gas particle neighbhour-finding routines that compute pressure, temperature and density gradients only search for gas particles with the same \texttt{ADMType}. We tested the hydrodynamic implementation by validating on differing initialisations of baryonic and ADM gases, making sure that both species were separate hydrodynamically. 

For cooling rates, we wrote separate routines containing all the ADM cooling functions in \citet{Rosenberg2017} with the correct scalings with ADM particle physics parameters. For each ADM cooling process listed in Sec.~\ref{sec:dark_cooling_rates}, we checked the output values from the functions as well as the \texttt{GIZMO} output during controlled hydrodynamic tests to verify the implementation and the ADM gas cooling scaling with varying ADM parameters according to Fig.~\ref{fig:cooling_rate_varying_params}. 

We also ensured that all essential baryonic feedback processes such as baryonic star formation criteria, baryonic feedback, etc.~were completely decoupled from the ADM gas. Moreover, the ADM gas has an entirely different clump-formation criteria as well as dark feedback physics (in this investigation, there is no dark feedback).

\clearpage
\section{Supplementary Figures and Tables}
\label{sec:supplementary_figures}

\setcounter{equation}{0}
\setcounter{figure}{0}
\setcounter{table}{0}
\renewcommand{\theequation}{E\arabic{equation}}
\renewcommand{\thefigure}{E\arabic{figure}}
\renewcommand{\thetable}{E\arabic{table}}

\hspace{-1.5cm}
\begin{table}[h]
\centering
\renewcommand{\arraystretch}{1.4}
\begin{tabular}{c|cc|cc|cccc|cccc}
  \Xhline{3\arrayrulewidth}
   & \multicolumn{2}{c|}{$\cdm$} & \multicolumn{2}{c|}{$\cdmnf$} & \multicolumn{4}{c|}{$\admone$} & \multicolumn{4}{c}{$\admtwo$}\\
 \cline{2-13} 
& \multicolumn{2}{c|}{\textbf{Baryons}}  & \multicolumn{2}{c|}{\textbf{Baryons}} & \multicolumn{2}{c|}{\textbf{Baryons}}        & \multicolumn{2}{c|}{\textbf{ADM}} & \multicolumn{2}{c|}{\textbf{Baryons} } & \multicolumn{2}{c}{\textbf{ADM}} \\ 
& Gas  & Stars  & Gas & Stars & Gas  & \multicolumn{1}{c|}{Stars} & Gas & Clumps & Gas & \multicolumn{1}{c|}{Stars} & Gas & Clumps      \\ 
\hline\hline
\multicolumn{1}{c|}{$\rhalf [\kpc]$}   & 3.83          & 1.83          & 1.42              & 1.58             & 3.35   & \multicolumn{1}{c|}{1.34}  & 1.64        & 0.63       & 1.19   & \multicolumn{1}{c|}{1.59}  & 2.91       & 0.78         \\ 
\multicolumn{1}{c|}{$\zhalf [\kpc]$}   & 0.039         & 0.28          & 0.035             & 0.22             & 0.022 & \multicolumn{1}{c|}{0.17}  & 0.014      & 0.14       & 0.022 & \multicolumn{1}{c|}{0.21}  & 0.015      & 0.18        \\ 
\multicolumn{1}{c|}{$\zninety [\kpc]$} & 0.12          & 0.99          & 0.119              & 1.64             & 0.087  & \multicolumn{1}{c|}{0.63}  & 0.045       & 0.75       & 0.086  & \multicolumn{1}{c|}{0.75}  & 0.035      & 0.71        \\ 
\multicolumn{1}{c|}{$\fthin$}          & 0.94             & 0.38          & 0.67                  & 0.20             & 0.97      & \multicolumn{1}{c|}{0.39}  & 0.86           & 0.19       & 0.81      & \multicolumn{1}{c|}{0.37}  & 0.99          & 0.21        \\ 
\multicolumn{1}{c|}{$\fthick$}         & 0.05             & 0.39          & 0.32                  & 0.53             & 0.03      & \multicolumn{1}{c|}{0.40}  & 0.14           & 0.57       & 0.18      & \multicolumn{1}{c|}{0.45}  & 0.01          & 0.52        \\ 
\multicolumn{1}{c|}{$\fspheroid$}         & 0.01             & 0.23          & 0.01                  & 0.27             & 0.00      & \multicolumn{1}{c|}{0.21}  & 0.00           & 0.24       & 0.01      & \multicolumn{1}{c|}{0.18}  & 0.00          & 0.27        \\ 
\multicolumn{1}{c|}{$\flatness$}       & 0.95          & 0.71          & 0.95               & 0.51             & 0.96   & \multicolumn{1}{c|}{0.71}  & 0.98        & 0.53       & 0.96   & \multicolumn{1}{c|}{0.68}  & 0.97       & 0.50        \\ 
  \Xhline{3\arrayrulewidth}
\end{tabular}
\caption{\label{tab:Morphology_data} A table of morphology metrics for the baryonic stars and ADM clumps in $\cdm$, $\cdmnf$, $\admone$, and $\admtwo$. $\rhalf,\, \zhalf, \, \zninety, \, \fthin,\, \fthick,$ and $\fspheroid$ are all defined in the main text.  $\flatness$ is the flatness parameter, with $\flatness \rightarrow 1$ approaching a thin-disk distribution and $\flatness \rightarrow 0$ approaching a spherical distribution. To obtain $\flatness$, we compute the moment of inertia tensor of the stars or ADM clumps in the central $10\kpc$ of the galaxy and compare the values to that of a uniform ellipsoid, obtaining its effective triaxial dimensions. We then repeat the process with particles within the derived ellipsoid boundaries until the boundary values converge to within $10\%$. Flatness is then defined as $\flatness = 1-c/a$, where $a$~($c$) is the final semi-major~(semi-minor) value of the iterative calculation.
}
\end{table}

\begin{figure}[h]
    \centering 
    \includegraphics[trim={1cm 2.5cm 0cm 2cm},clip,width=0.5\textwidth]{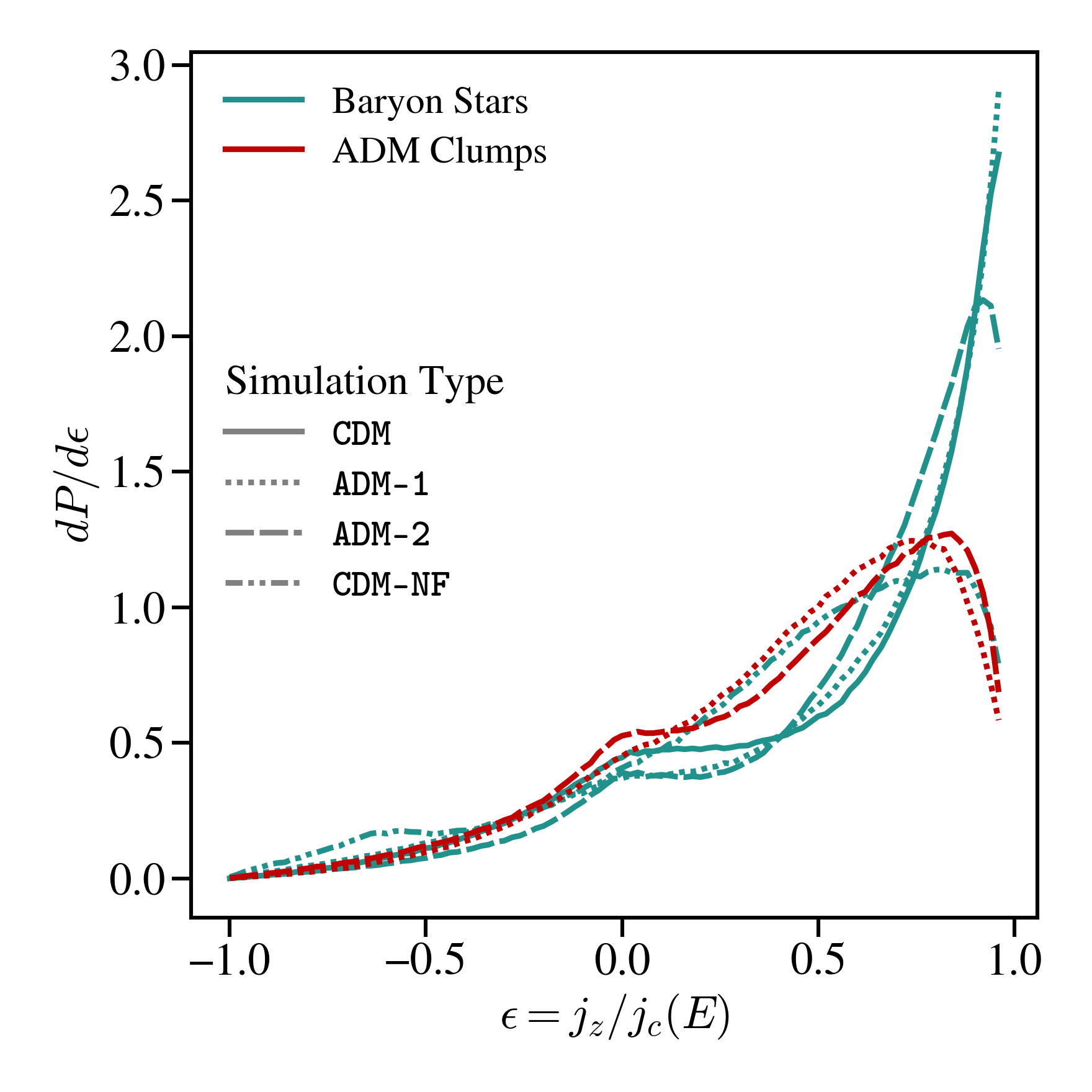} 
    \caption{\label{fig:orbital_circ} The orbital circularity probability density functions for baryonic stars and ADM clumps in $\cdm$, $\cdmnf$, $\admone$, and $\admtwo$ at $z\sim0$.  Orbital circularities in the range $\epsilon \geq 0.8$, $0.2 < \epsilon < 0.8$, and $\epsilon < 0.2$ resemble a thin disk, thick disk, and spheroid, respectively. Baryonic stars in the $\cdm$, $\admone$ and $\admtwo$ simulations have dominant thin disks, while those in $\cdmnf$ have more prominent thick disks. The same holds for feedback-less ADM clumps in $\admone$ and $\admtwo$.}
\end{figure}

\clearpage

\begin{figure*}[h]
    \raggedright
    \includegraphics[width=0.325\linewidth]{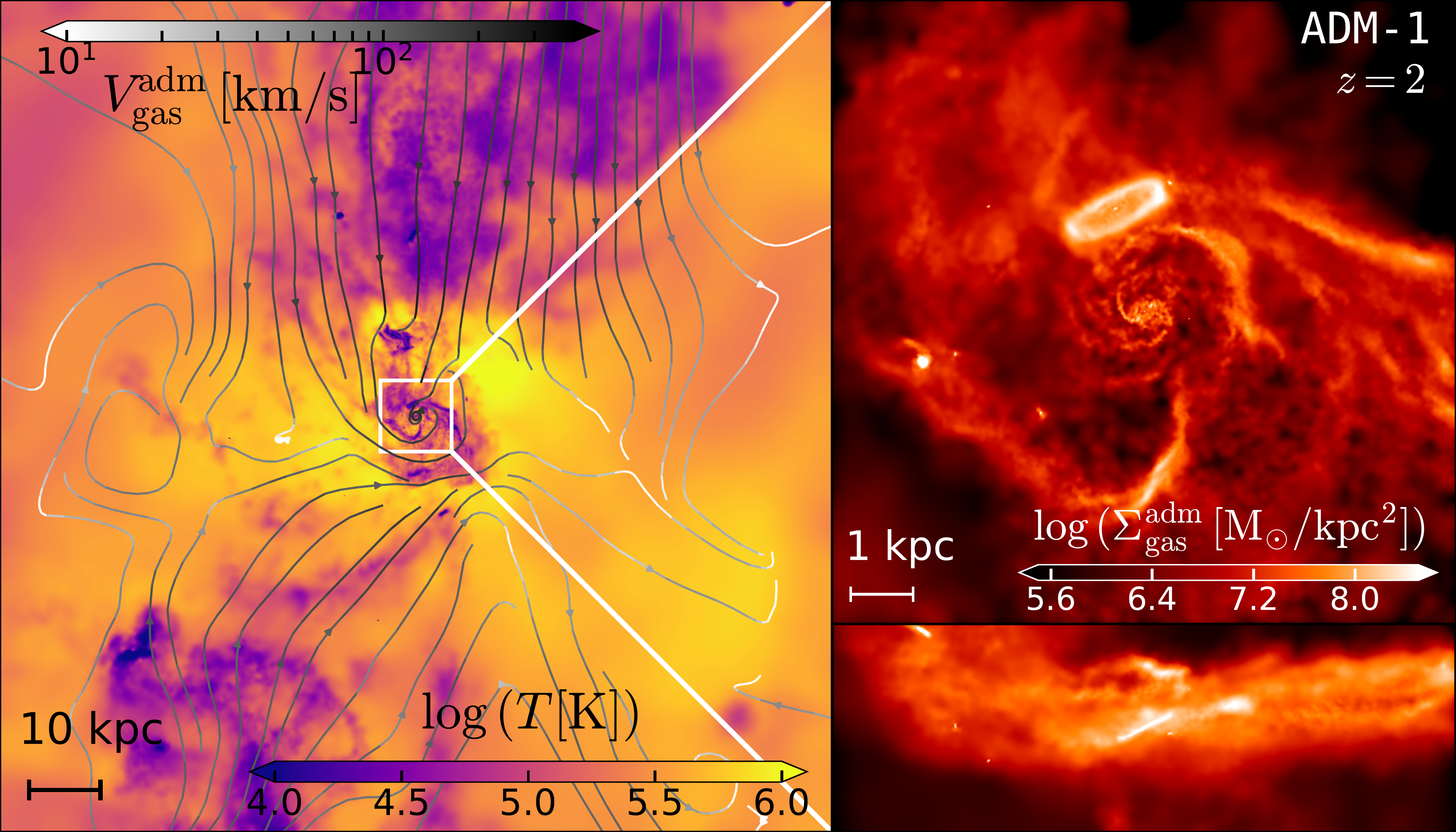}
    \includegraphics[width=0.325\linewidth]{figures/mosaics/mosaic_adm1pt35_adm_277.pdf}
    \includegraphics[width=0.325\linewidth]{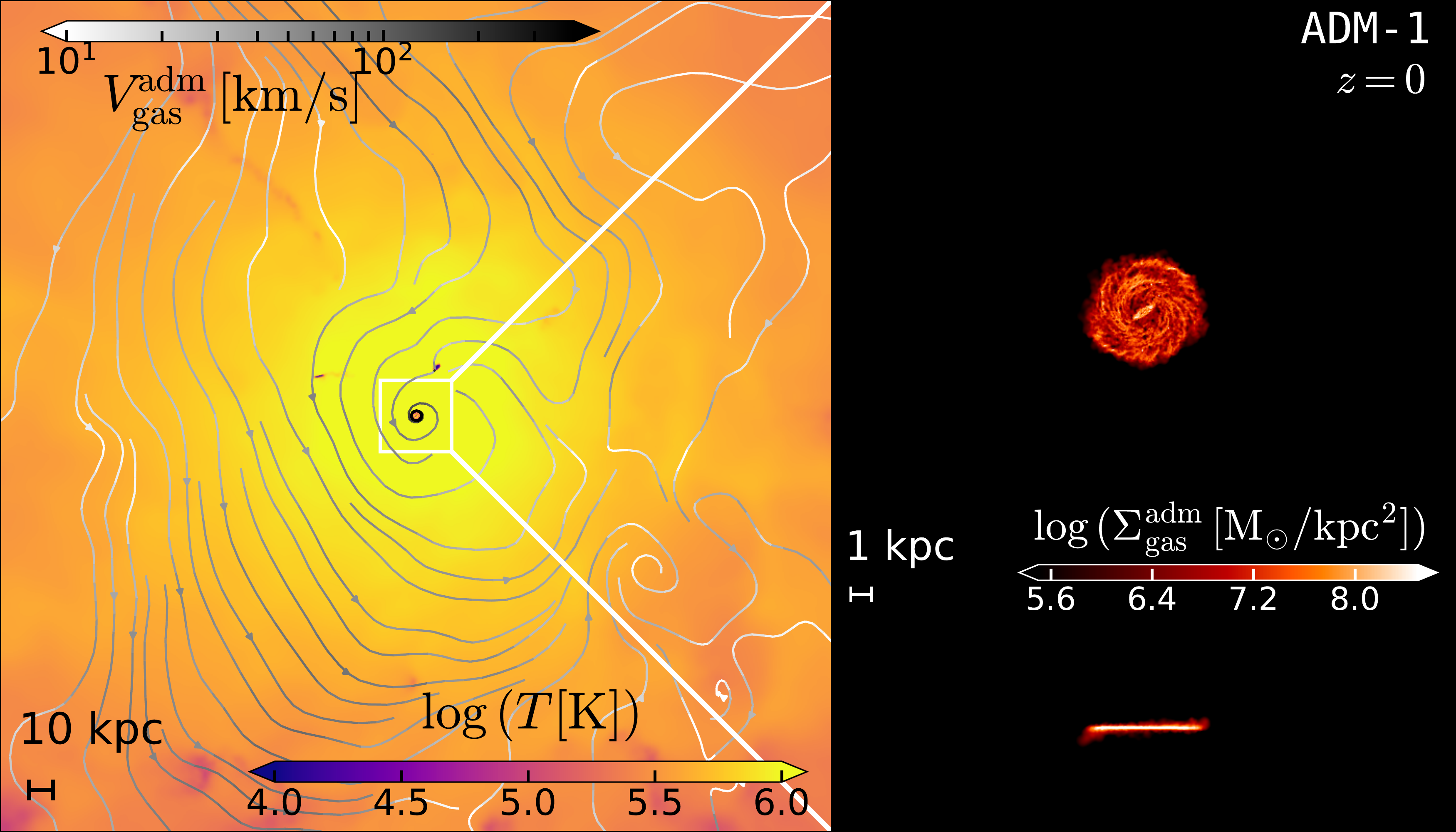}
    
    \includegraphics[width=0.325\linewidth]{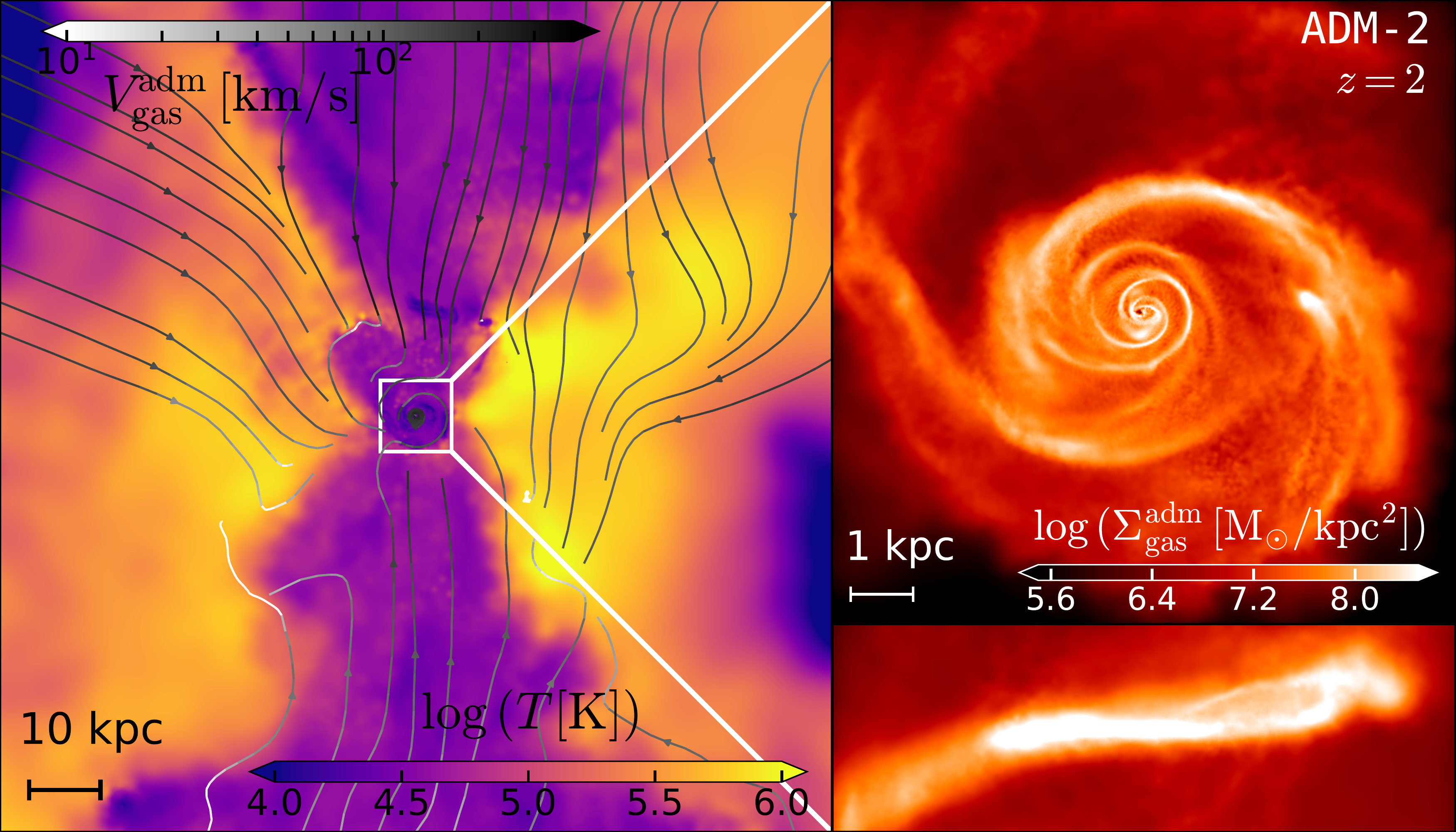}
    \includegraphics[width=0.325\linewidth]{figures/mosaics/mosaic_adm2pt5_adm_277.pdf}
    \includegraphics[width=0.325\linewidth]{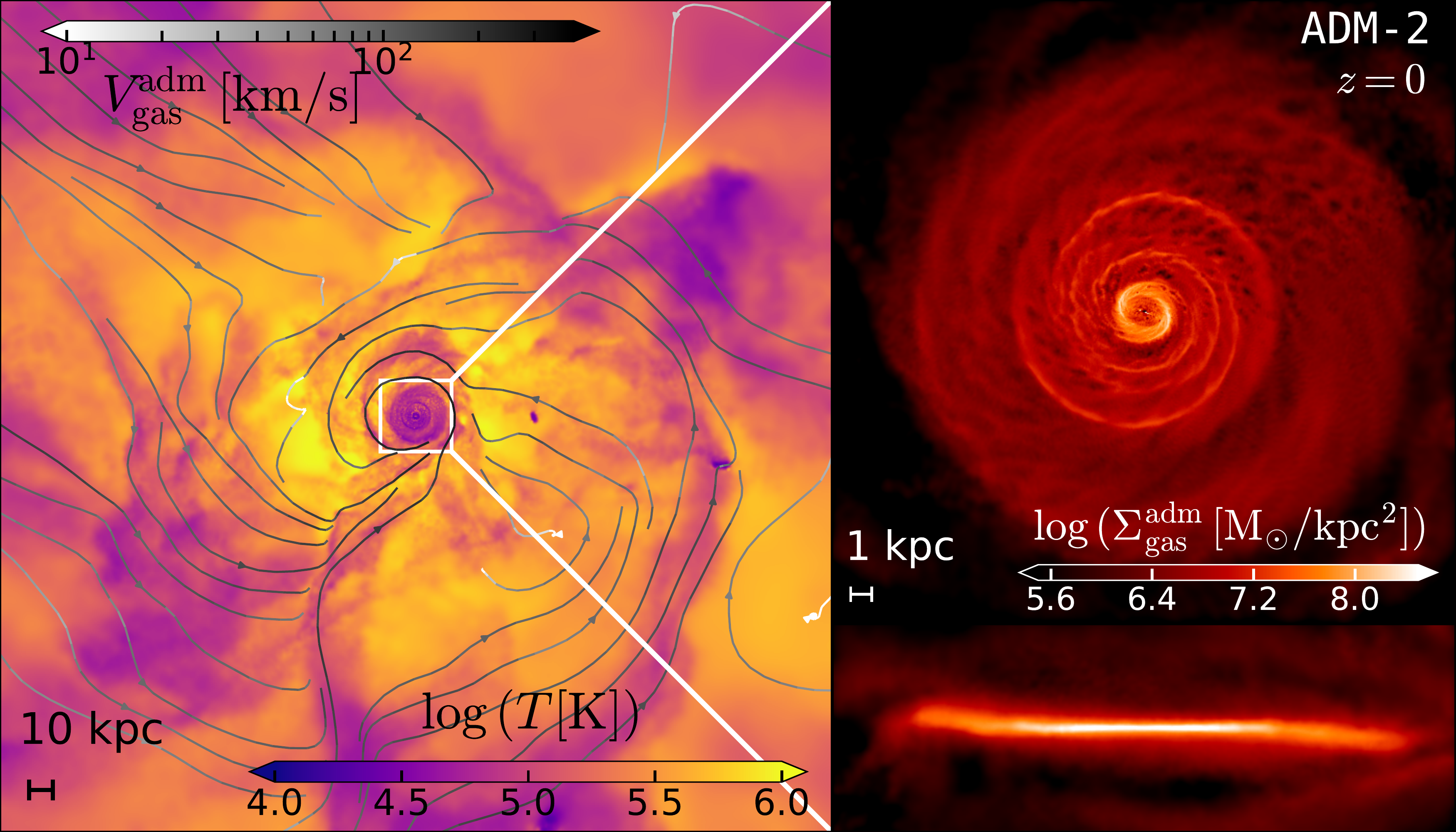}

    \includegraphics[width=0.325\linewidth]{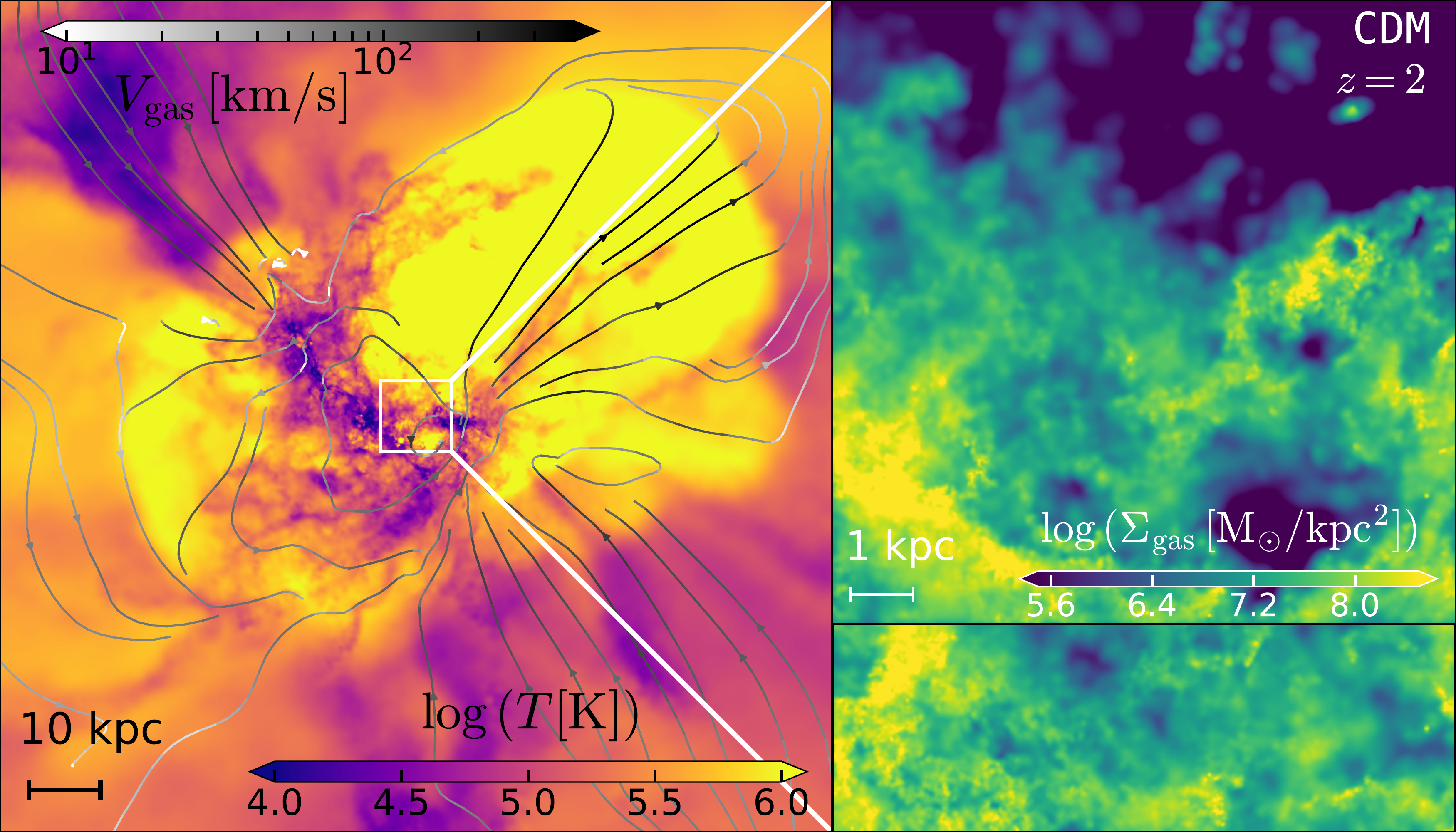}
    \includegraphics[width=0.325\linewidth]{figures/mosaics/mosaic_cdm_baryon_277.pdf}
    \includegraphics[width=0.325\linewidth]{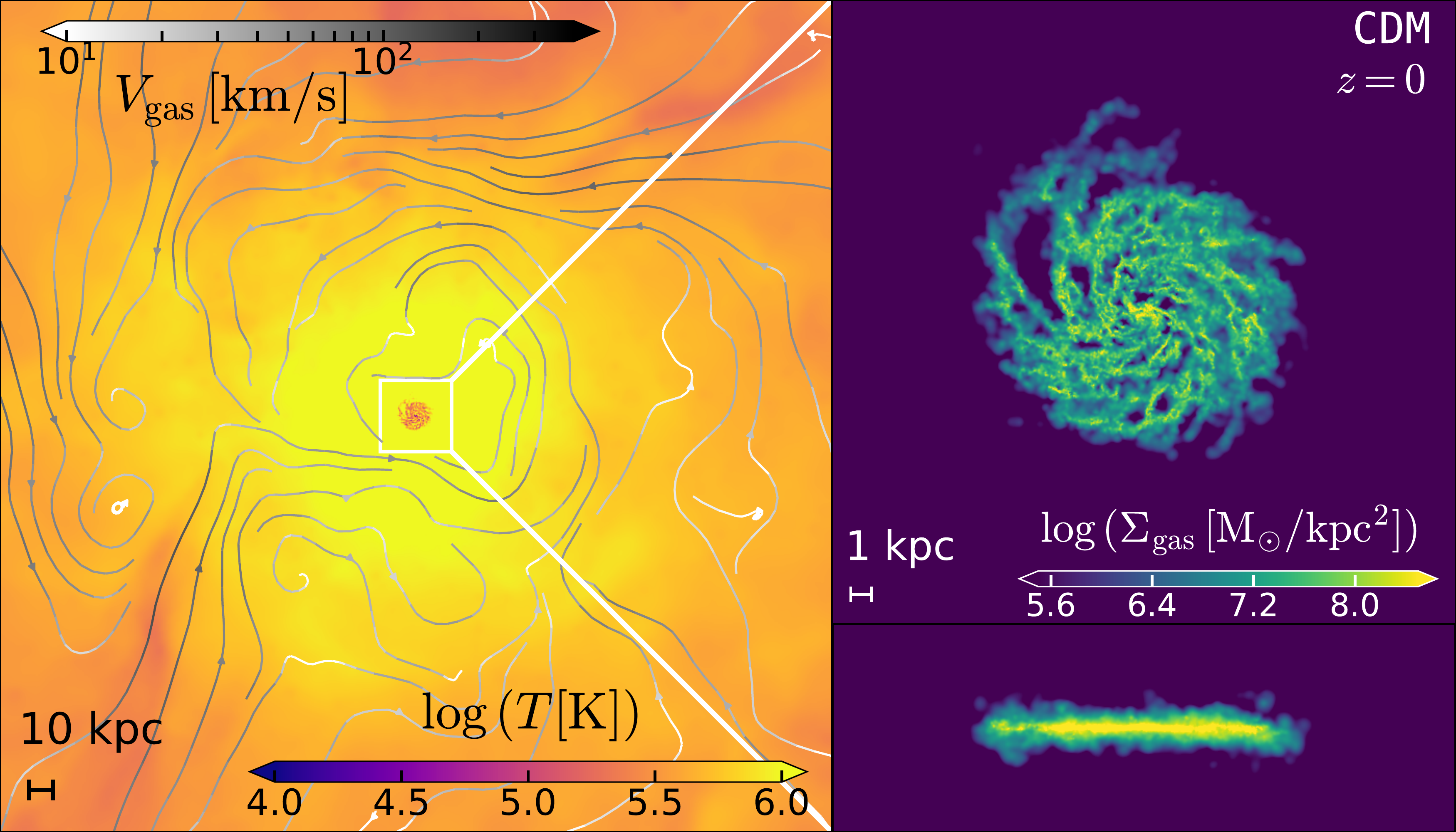}
    
    \includegraphics[width=0.325\linewidth]{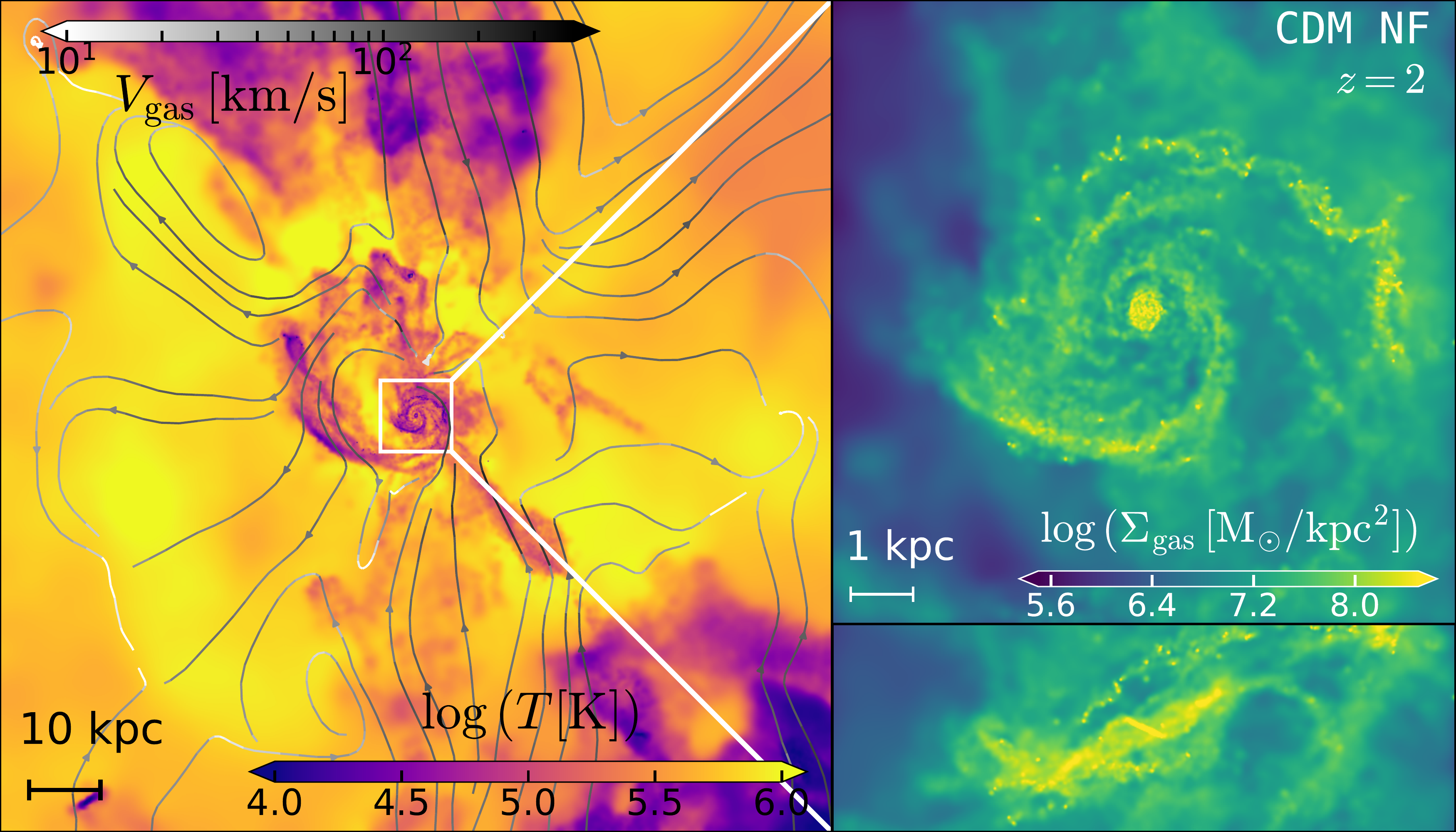}
    \includegraphics[width=0.325\linewidth]{figures/mosaics/mosaic_cdmnoFB_baryon_290.pdf}
    \includegraphics[width=0.325\linewidth]{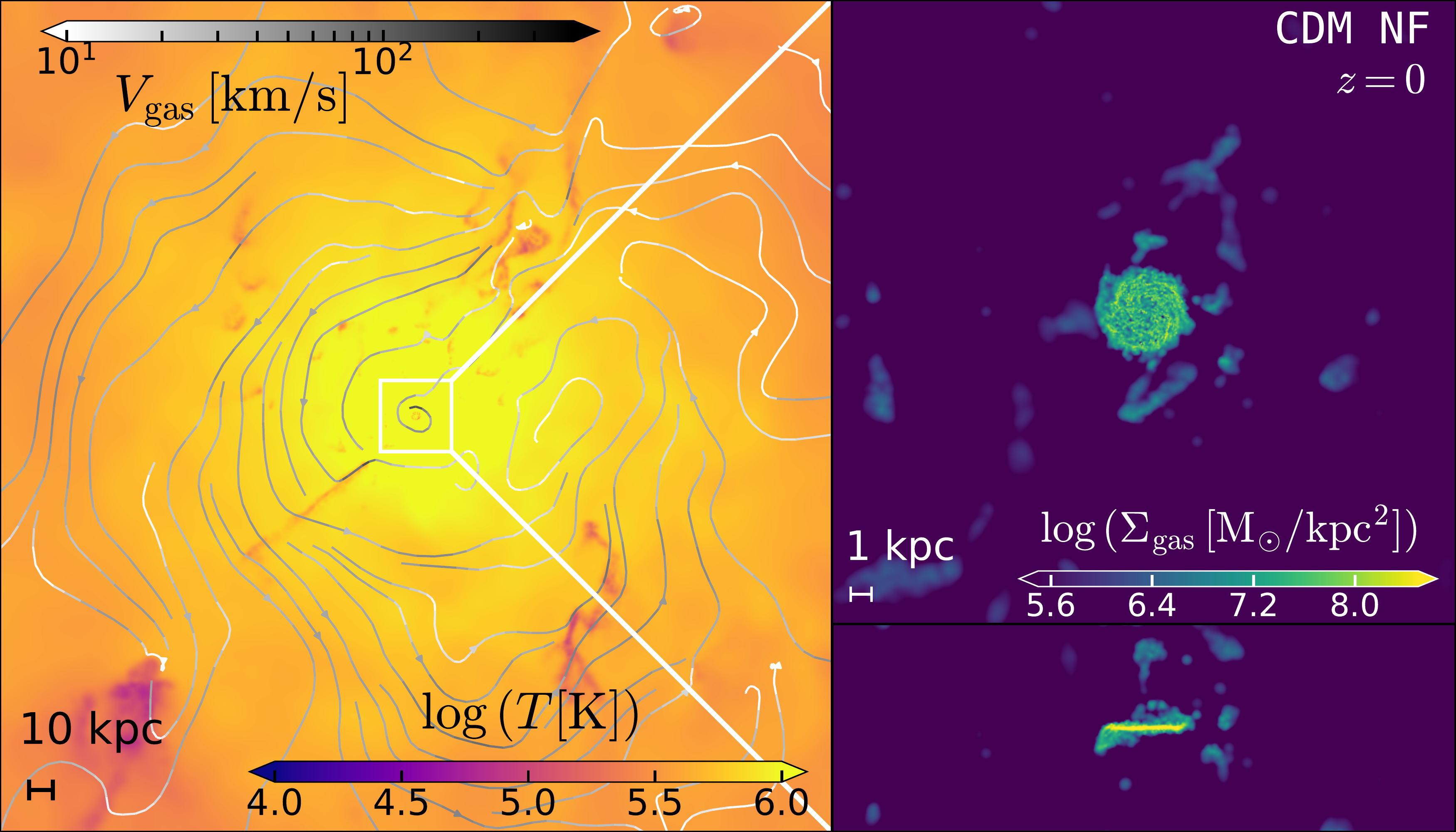}
    \caption{Same as Fig.~\ref{fig:baryon_gas}, except comparing the $z\sim2$, $z\sim1$, and $z\sim0$ distributions (from left to right). The comoving field-of-view is kept consistent across images at different redshifts. The gas temperature maps show that, across all models, the inner CGM transitions from an unstable state with supersonic cold gas inflows to a stable virialized state dominated by hot-mode accretion. The timing of this transition is influenced by the gravitational potential of the halo and cooling physics. The transition time is comparable for $\admone$, $\cdm$, and $\cdmnf$ simulations due to their comparable cooling functions. The transition appears to be delayed in $\admtwo$ due to its significantly higher cooling rate. A minor difference between CDM simulations (both $\cdm$ and $\cdmnf$) to $\admone$ is the multiphase structure of the CGM, which arises from helium and metal-cooling peaks in the cooling function, that leads to thermal instabilities. $\cdm$ displays a chaotic gas distribution near the galactic center at $z\sim2$ due to feedback, which  settles down to a disk by $z\leq 1$. In contrast, $\admone$, $\admtwo$, and $\cdmnf$ feature early-settled disks at $z\sim2$, which are then disturbed by galaxy mergers and cold filaments of accreted gas at around $z\sim1$. The final gas distribution in $\admone$ and $\cdmnf$ are more compact. However, $\admtwo$, which has strong cooling rates and late-time gas accretion, maintains an extended gas disk at $z\sim0$ similar to baryons in $\cdm$. Videos of the simulations are available \href{https://rb.gy/et2q0}{here}.}
    \label{fig:baryon_gas_mosaic}
\end{figure*}

\clearpage
\begin{figure*}[h]
    \centering 
    \includegraphics[trim={0cm 0cm 0cm 0cm},clip,width=1.0\textwidth]{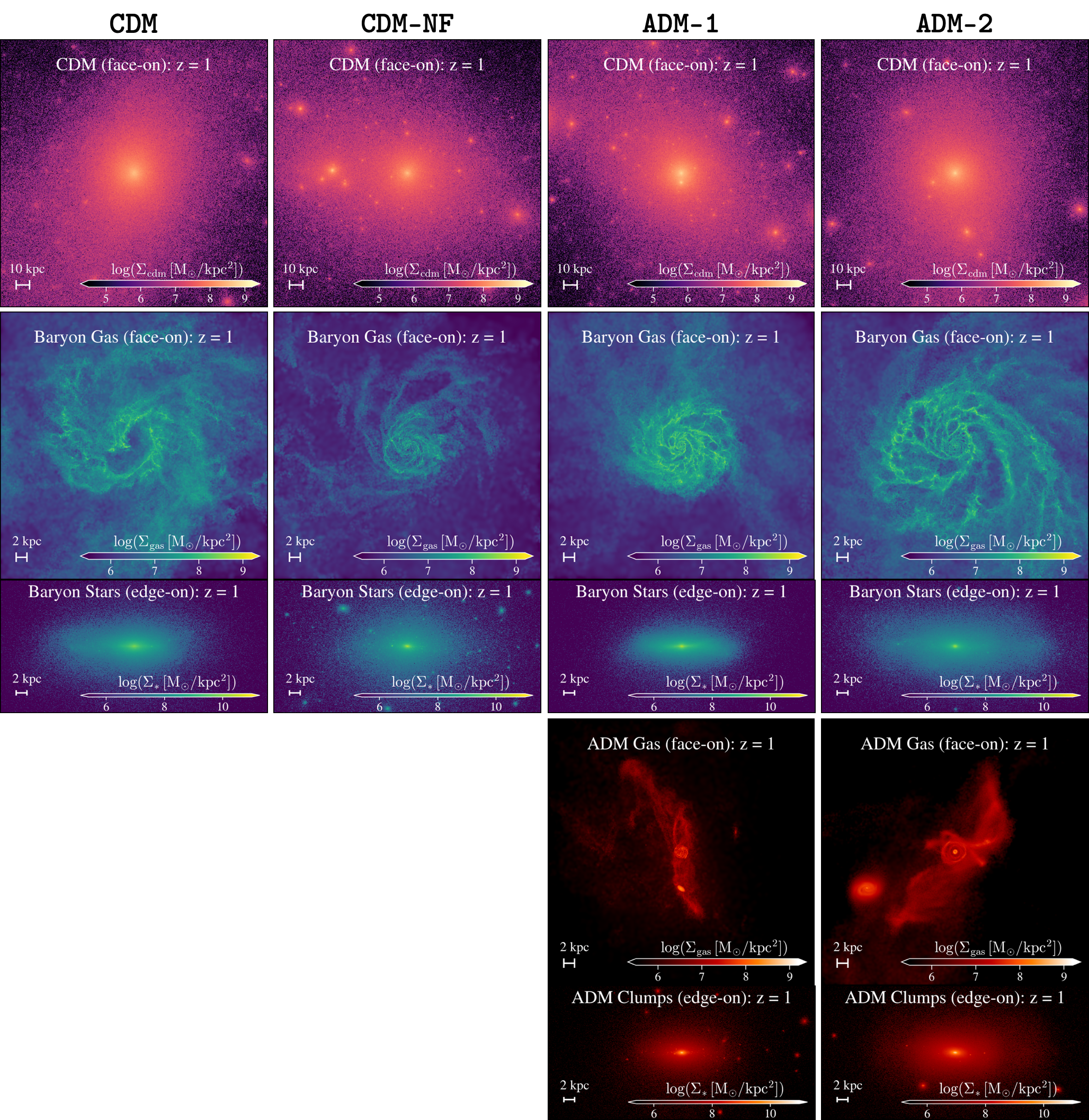}    \caption{\label{fig:intermediate_morphologies} Same as Fig.~\ref{fig:final_baryon_images}, except for $z\sim1$.  The top row also provides the density distributions for CDM. Videos of the simulations are available \href{https://rb.gy/et2q0}{here}.
} 
\end{figure*}

\clearpage
\begin{figure*}[h]
    \centering 
    \includegraphics[trim={0cm 0cm 0cm 0cm},clip,width=1.0\textwidth]{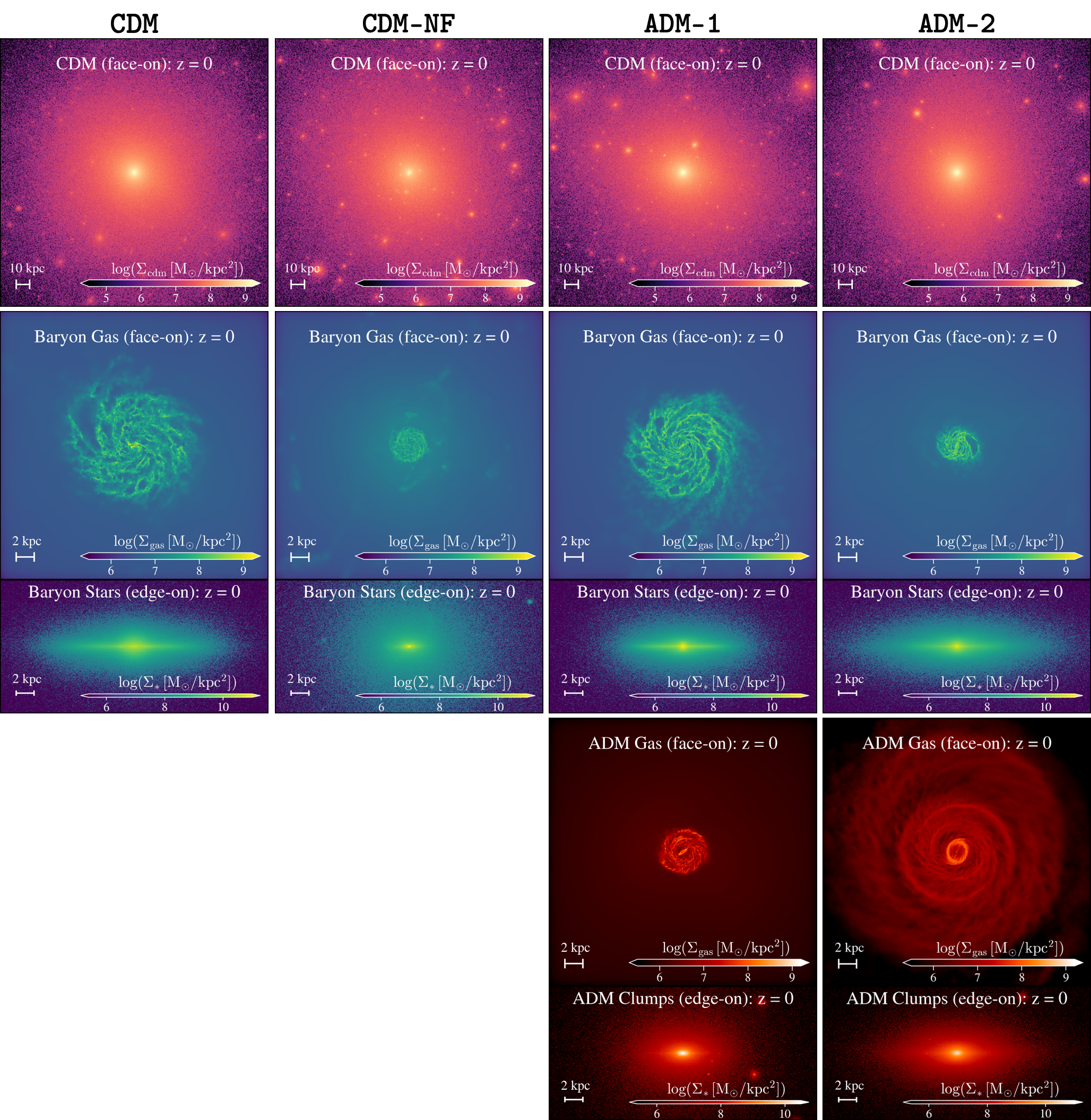} \caption{\label{fig:final_morphologies} Same as the $z\sim0$ density distributions in Fig.~\ref{fig:final_baryon_images}, except supplemented with the CDM densities in the top row. Videos of the simulations are available \href{https://rb.gy/et2q0}{here}.
}
\end{figure*}

\clearpage


\begin{figure}[h]
    \centering 
    \includegraphics[trim={0cm 0cm 0cm 0cm},clip,width=1.0\textwidth]{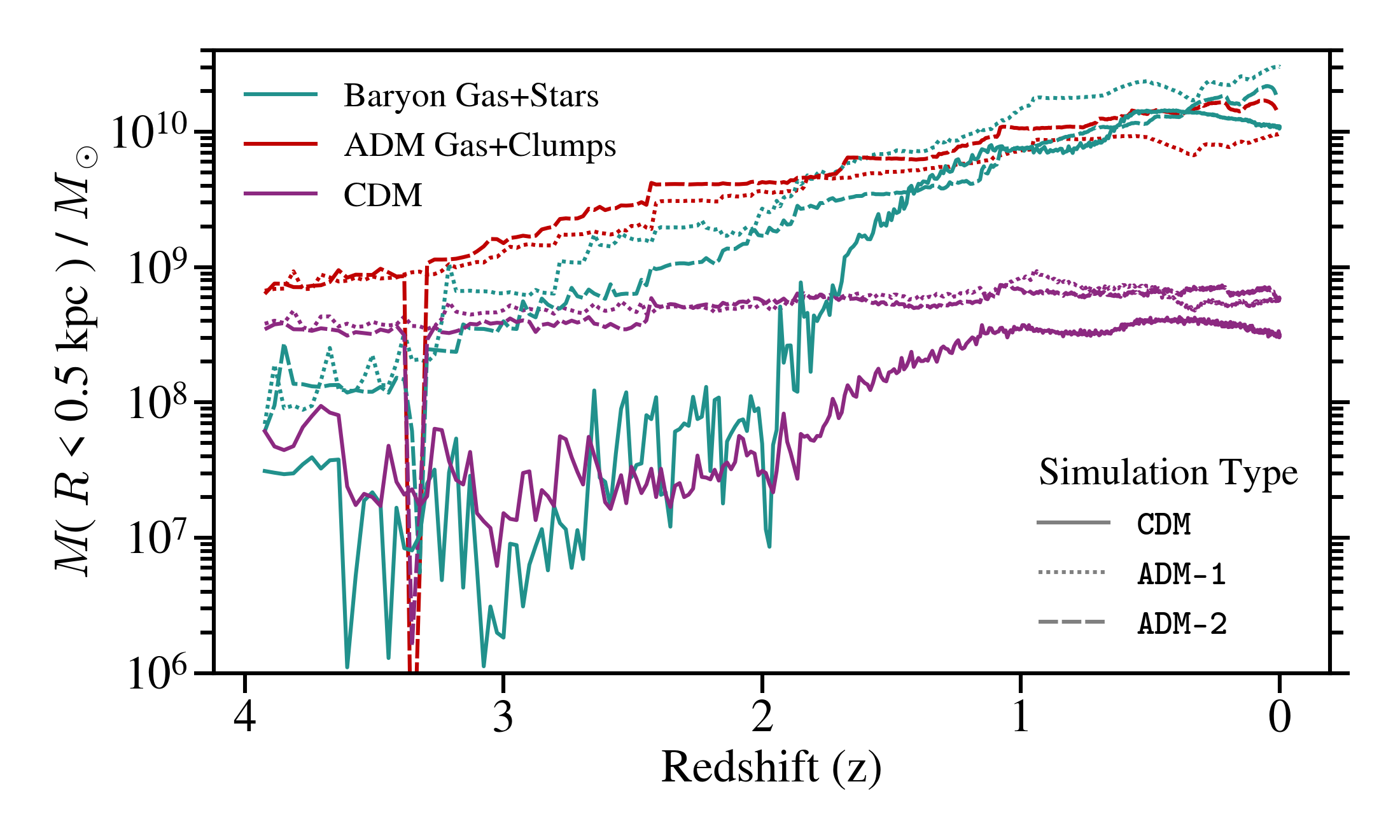}    \caption{\label{fig:central_density_evolution} A plot of the central mass contained in the inner 0.5$\kpc$ of each simulated galaxy, as a function of redshift. ADM dominates the inner galactic density at redshifts $z \gtrsim 3$ and enhances the central densities of baryons and CDM in both $\admone$ and $\admtwo$ compared to $\cdm$. Note that there is a sharp dip in the enclosed mass for $\admtwo$ species at $z\approx 3.3$ because a large merger temporarily perturbed the central galaxy and the center-finding algorithm locked onto a smaller mass galaxy.}
\end{figure}

\clearpage
\begin{figure}[h]
    \centering 
    \includegraphics[trim={3cm 0cm 3cm 2cm},clip,width=0.9\textwidth]{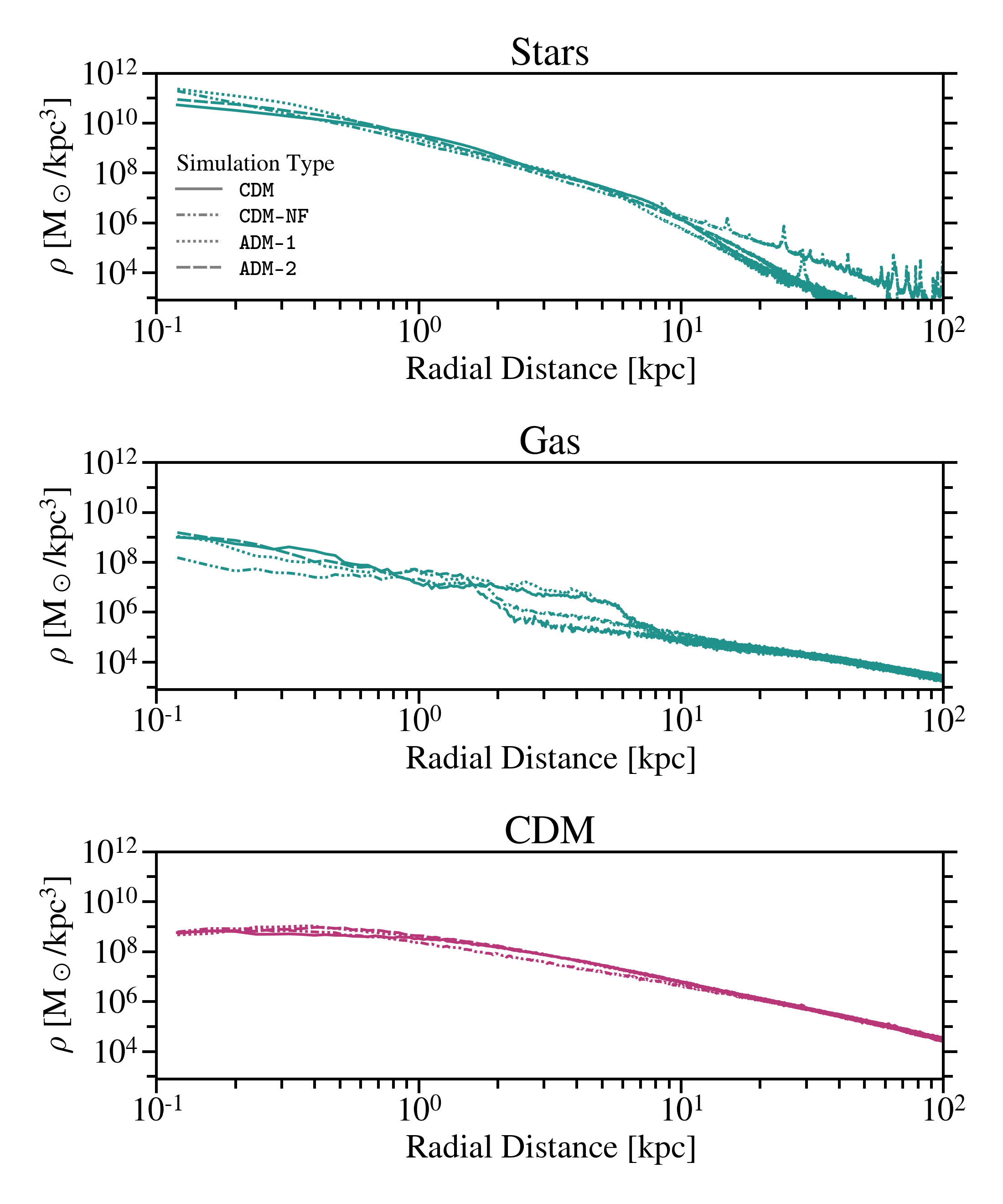} \caption{\label{fig:standard_density_profile} Density profiles for baryons and CDM.}
    \label{fig:density_profiles}
\end{figure}

\clearpage

\begin{figure}[h]
    \centering 
    \includegraphics[trim={0cm 0cm 0cm 0cm},clip,width=1.0\textwidth]{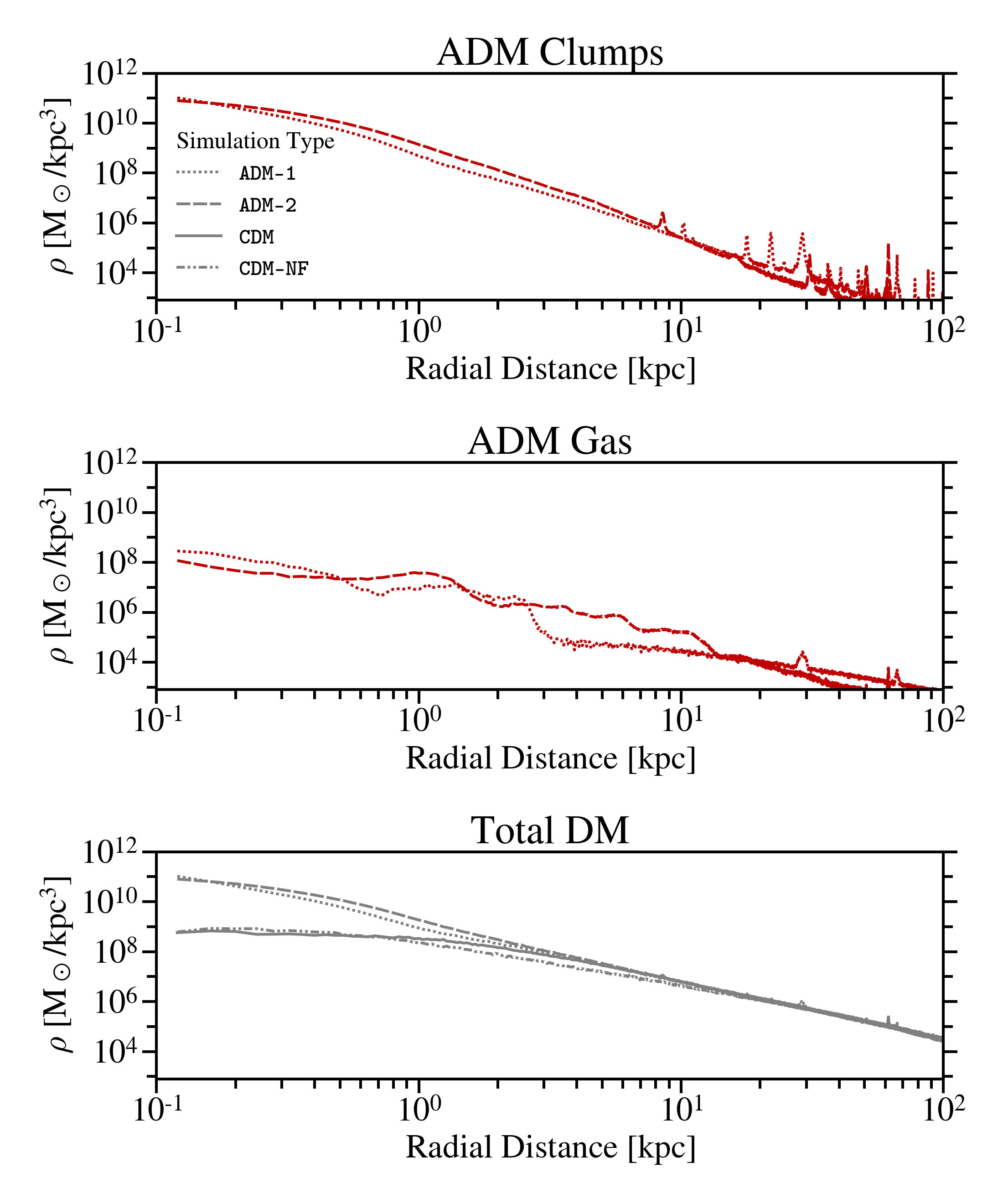} \caption{\label{fig:adm_density_profiles} Density profiles for Atomic Dark Matter species (upper two plots) and all dark matter (CDM for $\cdm$ and $\cdmnf$ and CDM+ADM for $\admone$ and $\admtwo$).}
\end{figure}

\begin{figure}[h]
    \centering 
    \includegraphics[trim={0cm 0cm 0cm 0cm},clip,width=1.0\textwidth]{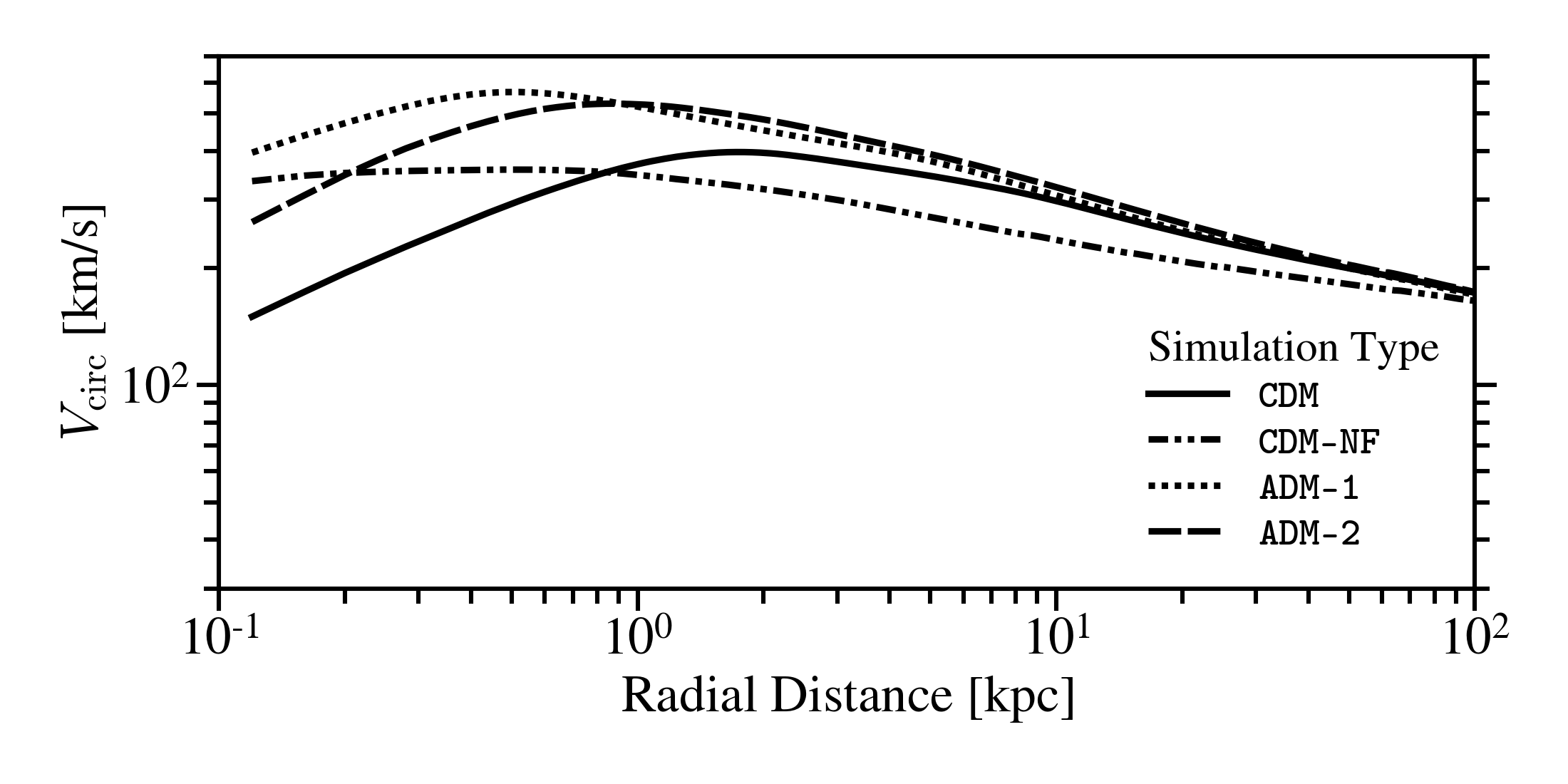} \caption{\label{fig:circular_v_profiles} Circular velocities ($\sqrt{G M(<r)/r}$) for each simulation.}
\end{figure}


\end{document}